\documentclass[11pt,a4paper]{article}

\usepackage{jheppub}
\usepackage{amsmath}
\usepackage{amssymb,amsthm}
\usepackage{mathtools}
\usepackage{mathrsfs}
\usepackage{relsize}
\usepackage{bm}
\allowdisplaybreaks[4]  
\usepackage{epsfig,graphics,graphicx}
\usepackage{color,xcolor}
\usepackage{subfigure}

\usepackage{array}
\usepackage{ragged2e}
\usepackage{lineno}
\usepackage{natbib}
\usepackage{hyperref}
\hypersetup{colorlinks=true,
            linkcolor=blue,
            anchorcolor=blue,
            citecolor=magenta,
            filecolor=blue,
            urlcolor=blue,
            bookmarksnumbered=true,
            pdfview=FitB
}
\usepackage{enumitem}
\usepackage{ulem}
\usepackage{listings}
\usepackage{soul}
\usepackage{cancel}

\usepackage{braket}

\colorlet{darkgreen}{green!50!black}
\colorlet{brightyellow}{yellow!75!red}
\colorlet{orange}{red!50!yellow}
\colorlet{darkgray}{gray!50!black}
\colorlet{darkred}{red!50!black}

\def\dd{{\mathrm{d}}}

\graphicspath{{./Images/}}

\makeatletter
\newcommand*{\transpose}{%
  {\mathpalette\@transpose{}}%
}
\newcommand*{\@transpose}[2]{%
  \raisebox{\depth}{$\m@th#1\intercal$}%
}
\makeatother

\title{Quantum entanglement between partons in a strongly coupled quantum field theory}

\author[1]{Wenyu~Zhang,}
\emailAdd{zwy\_2020@mail.ustc.edu.cn}
 
\author[2]{Wenyang~Qian,}
\emailAdd{qian.wenyang@usc.es}

\author[3,4,5]{Yiyu~Zhou,}
\emailAdd{yiyu.zhou@unito.it}

\author[1, 6, *]{Yang~Li,}
\emailAdd{leeyoung1987@ustc.edu.cn}
\note[*]{Corresponding author}

\author[1, 6, 7]{Qun~Wang}
\emailAdd{qunwang@ustc.edu.cn}

\affiliation[1]{Department of Modern Physics, University of Science \& Technology of China, Hefei 230026, China}
\affiliation[2]{Instituto Galego de Física de Altas Enerxías IGFAE, Universidade de Santiago de Compostela, E-15782 Galicia-Spain}
\affiliation[3]{Department of Physics, University of Turin, via Pietro Giuria 1, I-10125 Torino, Italy}
\affiliation[4]{INFN, Section of Turin, via Pietro Giuria 1, I-10125 Torino, Italy}
\affiliation[5]{Center for Frontiers in Nuclear Science, Stony Brook University, Stony Brook, NY 11794, USA}
\affiliation[6]{Anhui Center for Fundamental Sciences in Theoretical Physics, Hefei, 230026, China}
\affiliation[7]{School of Mechanics and Physics, Anhui University of Science and Technology, Huainan, Anhui 232001, China}


\arxivnumber{}

\date{\today}

\abstract{
We perform a first-principles, non-perturbative investigation of quantum entanglement between partonic constituents in a strongly coupled 3+1-dimensional scalar Yukawa theory, using light-front Hamiltonian methods with controlled Fock-space truncations. By explicitly constructing reduced density matrices for (mock) nucleon, pion, and anti-nucleon subsystems from light-front wave functions, we compute key entanglement witnesses, including von Neumann entropy, mutual information, and linear entropy, in both quenched (no sea pairs) and unquenched frameworks. 
We find that the entanglement entropy is closely related to the Shannon entropy of the transverse momentum dependent distribution, establishing a link between quantum information and parton structure. In contrast, the unquenched theory reveals genuinely non-classical correlations: the entanglement entropy cannot be reduced to any Shannon entropy of normalized parton distributions, demonstrating that the full hadronic wave function encodes quantum information beyond classical probabilities.  
Our findings highlight the role of entanglement as a fundamental probe of non-perturbative dynamics in relativistic quantum field theory and lay the groundwork for extending these concepts to QCD and future collider phenomenology.
}

\begin{document}
\maketitle

\section{Introduction}

Modern high-energy collider experiments are fundamentally grounded in the framework of collinear factorization, which separates the dynamics of hard scattering from the non-perturbative structure of hadrons \cite{Gross:2022hyw}. This paradigm is encapsulated in the factorized cross section
\begin{equation}
\sigma_{ab} = \int \dd x_a \, \dd x_b \, \hat{\sigma}_{ij}(x_a,x_b; \mu^2_F) \, f_{i/a}(x_a;\mu^2_F) \, f_{j/b}(x_b;\mu^2_F),
\end{equation}
where $\hat{\sigma}_{ij}$ denotes the perturbatively calculable short-distance partonic cross section, and $f_{i/a}(x;\mu^2_F)$ is the parton distribution function (PDF) encoding the probability of finding a parton of flavor $i$ carrying a longitudinal momentum fraction $x$ inside hadron $a$ at the factorization scale $\mu_F$. Decades of experimental effort at facilities such as HERA, the Tevatron, and the LHC have enabled the extraction of a comprehensive set of PDFs across a wide kinematic range, providing a detailed empirical portrait of hadronic structure in terms of quarks and gluons \cite{ParticleDataGroup:2024cfk}.

From an information-theoretic perspective, the PDFs can be treated as classical probability distributions over the momentum fraction $x$. Consequently, one may compute their Shannon entropy,
\begin{equation}
H(f) =\log K -\int_0^1 \dd x \, f(x) \log f(x),
\end{equation}
where $K$ is the resolution of $x$, ensuring a positive entropy. However, this observation leads to a conceptual paradox: the proton itself is a pure quantum state in the Hilbert space of QCD, and as such, its entropy must vanish identically. The non-zero Shannon entropy derived from PDFs therefore signals a loss of quantum information in the current experimental measurements of the proton structure. As emphasized by Kharzeev and others~\cite{Kharzeev:2017qzs, Kharzeev:2021nzh}, this implies that a vast reservoir of quantum information -- essential for understanding non-perturbative phenomena such as confinement, chiral symmetry breaking, and emergent hadronic mass -- is rendered inaccessible within the standard collinear factorization picture.

From a quantum-information standpoint, the Shannon entropy of a PDF can be interpreted as quantifying the entanglement between the observed parton (the ``system'') and the unobserved remainder of the hadron (the ``environment''), which includes soft gluons, sea quarks, and other degrees of freedom (d.o.f.) integrated out during the factorization procedure. This perspective shifts the focus from classical ignorance to genuine quantum correlations, motivating a rigorous investigation of parton entanglement as a window into the quantum structure of hadrons.

Entanglement in quantum field theories (QFTs) has been studied extensively, most notably through the entanglement entropy between spatial regions governed by the celebrated area law \cite{Nishioka:2018khk, Witten:2018zxz}. However, such geometric entanglement is ill-suited to collider physics, where measurements are performed on particles rather than on field configurations within spatial subregions. Alternative approaches have employed phenomenological models to explore entanglement in the partonic structure \cite{Liu:2018gae, CarrascoMillan:2018ufj, Beane:2019loz, Feal:2020myr, Liu:2022ohy, Liu:2022hto, Liu:2022qqf, Ehlers:2022oke, Benito-Calvino:2022kqa, Liu:2023zno, Grieninger:2023ufa, Florio:2023mzk, Ozzello:2025tfu, Florio:2025xup, Galvez-Viruet:2025rmy}, particularly in the valence-dominated large-$x$ regime \cite{Dumitru:2022tud, Dumitru:2023fih, Dumitru:2023qee, Dosch:2023bxj, Qian:2024fqf, Dumitru:2025bib, Kolbusz:2025vhh}, or the dense gluon-dominated small-$x$ regime \cite{Hagiwara:2017uaz, Kovner:2018rbf, Peschanski:2019yah, Armesto:2019mna, Tu:2019ouv, Duan:2020jkz, Gotsman:2020bjc, Ramos:2020kaj, Kharzeev:2021yyf, Zhang:2021hra, Duan:2023zke, Gursoy:2023hge, Datta:2024hpn, Levin:2024wtl, Ramos:2025tge, Grieninger:2025wxg, Kutak:2025awi, Sheikhi:2025hep}, often invoking principles such as maximum entropy \cite{Cervera-Lierta:2017tdt, Beane:2018oxh, Hentschinski:2022rsa, Asadi:2022vbl, Asadi:2023bat, Hentschinski:2023izh, Hatta:2024lbw}. While insightful, these efforts typically lack a direct connection to the non-perturbative formalism of QFT, failing to provide a first-principles derivation of entanglement from hadronic wave functions.

To bridge this gap, we adopt a non-perturbative approach based on light-front quantization of a scalar quantum field theory  (\ref{eqn:scalar_yukawa_model}) in $3+1$ dimensions.
The classical Lagrangian density of this model reads \cite{Li:2014kfa, Li:2015iaw, Li:2015zah, Karmanov:2016yzu, Hiller:2016itl}, 
\begin{equation}\label{eqn:scalar_yukawa_model}
 \mathscr L = \partial_\mu N^\dagger \partial^\mu N - m^2 N^\dagger N \\
 + \frac{1}{2}\partial_\mu \pi \partial^\mu \pi - \frac{1}{2}\mu^2 \pi^2 + g N^\dagger N \pi\,.
 \end{equation}
  This Lagrangian describes a complex scalar field $N$ interacting with a real scalar field $\pi$ via a Yukawa type coupling. It serves as a rudimentary model for the nucleon-pion interaction. Therefore, we will refer to the complex scalar field $N$ as the (mock) nucleon field and tentatively assign the mass of the nucleon $m = 0.94\,\mathrm{GeV}$ to its physical mass. Similarly, the real scalar field $\pi$ will be referred to as the (mock) pion field, and tentatively assigned the pion mass $\mu = 0.14\,\mathrm{GeV}$.  
  
Our approach offers several clear advantages for clarifying quantum entanglement between partons. First, light-front quantization provides a formal field-theoretical definition of the parton picture that extends the collinear partons used in the factorization formula \cite{Brodsky:1997de, Carbonell:1998rj, Bakker:2013cea, Brodsky:2022fqy}. In this augmented parton picture, each parton has both the longitudinal momentum $xP^+$, and the transverse momentum $\vec p_\perp$. 
Additionally, the light-front Hamiltonian formalism possesses a Galilean subgroup of the Poincar\'e group, enabling a clean separation between center-of-mass and intrinsic dynamics, which is useful for obtaining analytically tractable expressions for the entanglement entropy \cite{Heinzl:2000ht, Miller:2000kv, Vary:2009gt}. 
Furthermore, it is found that light-front vacuum is simple (though not trivial), and requires fewer quantum resources to simulate \cite{Alterman:2025prb}. 
Finally, in the scalar theory, the absence of gauge symmetry sidesteps complications associated with ghost fields and gauge redundancy, while still retaining the essential features of UV and IR divergences that mirror those in QCD. Crucially, absent the soft collinear singularities inherent to gauge theories, the partonic Fock expansion remains well-defined even at arbitrarily small distance scales, allowing us to resolve the quantum entanglement among partons with high accuracy \cite{Li:2014kfa, Li:2015iaw, Duan:2024dhy}.

In this paper, we leverage these properties to compute the quantum entanglement between partonic constituents directly from the light-front wave function (LFWFs) of a composite particle, obtained by solving a strongly coupled scalar theory in 3+1D. The remainder of the article is organized as follows. 
Section~\ref{sec:entanglement_witness} provides a rudimentary introduction to entanglement witnesses and their applications in QFTs. 
Section~\ref{sec:lf_wf} presents the non-perturbative solution of the model, the resulting LFWFs, and the derivation of parton distributions from them. 
Section~\ref{sec:density_matrix} discusses the construction of reduced density matrices, which are then employed in Secs.~\ref{sec:entanglement_in_quenched_theory} and~\ref{sec:entanglement_in_unquenched_theory} to investigate quantum entanglement in the quenched and unquenched theories, respectively. 
Further discussions, including scale dependence, rapidity dependence, and maximal entropy, are provided in Sec.~\ref{sec:discussions}. 
Finally, Sec.~\ref{sec:summary} offers a summary of our findings and an outlook on potential extensions to QCD.

\section{Entanglement witnesses}\label{sec:entanglement_witness}

In quantum mechanics, the state of a composite system $AB$ is described by a density operator $\rho_{AB}$ acting on the tensor product Hilbert space $\mathcal{H}_A \otimes \mathcal{H}_B$. A pure state $\ket{\Psi}_{AB}$ is said to be {separable} (i.e., unentangled) if it can be written as a product state $\ket{\psi}_A \otimes \ket{\phi}_B$. Otherwise, it is {entangled}. For mixed states, separability is defined more subtly: $\rho_{AB}$ is separable if it can be expressed as a convex combination of product states,
\[
\rho_{AB} = \sum_k p_k \, \rho_A^{(k)} \otimes \rho_B^{(k)}, \quad p_k \geq 0, \quad \sum_k p_k = 1.
\]
Any state that cannot be written in this form is entangled. 
Detecting entanglement in practice is nontrivial, especially when the full density matrix is unknown. An {entanglement witness} is an observable $\mathcal{W}$ such that $\operatorname{Tr}(\mathcal{W} \rho_{\text{sep}}) \geq 0$ for all separable states $\rho_{\text{sep}}$, but $\operatorname{Tr}(\mathcal{W} \rho_{\text{ent}}) < 0$ for at least one entangled state $\rho_{\text{ent}}$. See Refs.~\cite{Horodecki:2009zz, Guhne:2008qic} for review.  

In the context of high-energy physics, where direct access to the full quantum state is limited, entanglement witnesses serve as indispensable tools for diagnosing quantum correlations among subsystems such as partons inside a hadron \cite{Cheng:2025zaw, Afik:2025ejh, Qi:2025onf}.
While such operators are powerful in principle, constructing them requires detailed knowledge of the state space and is often impractical in QFT. Consequently, people frequently rely on {entanglement measures} -- quantities derived from the quantum state that are zero for separable states and positive for entangled ones. These serve as operational entanglement witnesses in theoretical and numerical studies.

For a pure bipartite state $\ket{\Psi}_{AB}$, the canonical measure of entanglement is the {entanglement entropy}, defined as the von Neumann entropy $S_\text{vN}$ of the reduced density matrix of either subsystem:
\begin{equation}
S_A \equiv S_\text{vN}(\rho_A) = -\operatorname{Tr}_A \left( \rho_A \log \rho_A \right), \quad \text{where} \quad \rho_A = \operatorname{Tr}_B \left( \ket{\Psi}\bra{\Psi} \right).
\end{equation}
By the Schmidt decomposition, $\ket{\Psi}_{AB} = \sum_n \sqrt{\lambda_n} \ket{n}_A \otimes \ket{n}_B$, with $\lambda_n \geq 0$ and $\sum_n \lambda_n = 1$, the entanglement entropy becomes
\begin{equation}
S_A = -\sum_n \lambda_n \log \lambda_n,
\end{equation}
which vanishes if and only if the state is a product state (i.e., only one $\lambda_n = 1$).
A closely related quantity is the Rényi entropy:
\begin{equation}
S_A^{(n)} \equiv S_\text{Ry}(\rho_A) = \frac{1}{1 - n} \log \operatorname{Tr} \left( \rho_A^n \right), \quad n > 0,\, n \neq 1.
\end{equation}
The limit $n \to 1$ recovers the von Neumann entropy. The second R\'enyi entropy ($n=2$) is particularly useful because $\operatorname{Tr}(\rho_A^2)$ can be measured experimentally \cite{Islam:2015mom} via quantum interference or computed efficiently in lattice simulations \cite{Hastings:2010zka}. The related {linear entropy},
\begin{equation}
S_{L}(\rho_A) = 1 - \operatorname{Tr}(\rho_A^2),
\end{equation}
provides a lower bound on the von Neumann entropy and is often used as a proxy in numerical work.
Beyond entropy-based quantifiers, a rich variety of entanglement witnesses has been developed to address different operational scenarios and mathematical structures. Examples include quantum negativity, mutual information, entanglement of formation \cite{Hill:1997pfa, Wootters:1997id} and local information \cite{Artiaco:2024noa, Barata:2025rjb, Artiaco:2025qqq} etc. 
In this work, we focus on the entanglement entropy as our primary witness. Its interpretation as the information loss due to tracing out unobserved partons aligns naturally with the factorization paradigm of perturbative QCD, and, crucially, it can be computed exactly from the LFWFs in non-perturbative models. 

In QFT, the Hilbert space is associated with field configurations over all space, and a natural way to define subsystems is by partitioning space into a region $A$ and its complement $\bar{A}$. The reduced density matrix $\rho_A = \operatorname{Tr}_{\bar{A}} \ket{\Psi}\bra{\Psi}$ is then obtained by tracing over field d.o.f.~in $\bar{A}$. This construction underlies the most extensively studied form of entanglement in QFT.
For gapped systems in $d+1$ spacetime dimensions, the entanglement entropy obeys the {area law} \cite{Eisert:2008ur}:
\begin{equation}
S_A \sim \mathcal{A}(\partial A) / \epsilon^{d-1},
\end{equation}
where $\mathcal{A}(\partial A)$ is the area of the boundary of region $A$, and $\epsilon$ is a UV cutoff. 
In contrast, conformal field theories (CFTs) in $1+1$ dimensions exhibit a logarithmic divergence,
\begin{equation}
S_A = \frac{c}{3} \log \left( \frac{L}{\pi a} \sin \frac{\pi \ell}{L} \right),
\end{equation}
where $c$ is the central charge, $\ell$ is the length of interval $A$, $L$ the system size, and $a$ a short-distance regulator. 
These results are foundational but rely on a geometric partition of space, which does not correspond to any measurement performed in collider experiments.
Recent efforts have shifted toward defining entanglement in momentum space between particle species. In this context, the parton distribution functions extracted from deep inelastic scattering can be interpreted as diagonal elements of a reduced density matrix in the longitudinal momentum basis \cite{Kharzeev:2021nzh, Hentschinski:2024gaa}. The associated Shannon entropy then serves as a coarse-grained proxy for the underlying quantum entanglement between the struck parton and the remnant hadronic system.

\section{Light-front wave function representation}\label{sec:lf_wf}

As discussed above, to investigate partonic entanglement in a controlled, non-perturbative framework, we employ a simplified yet physically rich model: the (3+1)D scalar Yukawa theory Eq.~(\ref{eqn:scalar_yukawa_model}). 
Physically, the theory describes a pointlike nucleon dressed by a cloud of virtual pions. The probability of finding the system in its bare, pointlike configuration is encoded in the field-strength renormalization constant $Z$, which satisfies $0 < Z \le 1$. 
 
 The coupling $g$ in this model is dimensionful, rendering the theory super-renormalizable. Despite this,  UV divergences appear at one-loop order in radiative corrections. To handle these, we adopt the Pauli-Villars (PV) regularization scheme \cite{Brodsky:1998hs}. After renormalization, the theory remains finite even in the limit where the PV regulator $\mu_{\text{PV}}$ is removed. For clarity, we suppress the explicit dependence on the PV regulators in the main text and relegate the fully regularized expression to Appendix~\ref{sec:PVreg}. 
  It is convenient to introduce a dimensionless coupling 
  \begin{equation}
  \alpha = g^2/(16\pi m^2),
  \end{equation}
  which corresponds to the strength of the tree-level Yukawa potential, 
 \begin{equation}
 V(r) = - \frac{\alpha}{r}e^{-\mu r}.
 \end{equation}

We quantize the theory on the light front $x^+ = 0$, where the light-front coordinates are defined as
\begin{equation}
x^\pm = x^0 \pm x^3, \quad  \vec x_\perp = (x^1, x^2).
\end{equation}
Physical states are eigenstates of the light-front Hamiltonian $\mathcal P^-$, satisfying the light-front Schrödinger equation, 
\begin{equation}\label{eqn:LF_Schrödinger_equation}
    \mathcal P^- |\psi(p)\rangle = \frac{\vec p^2_\perp+M^2}{p^+}|\psi(p)\rangle.
\end{equation}
In Fock space, this eigenvalue problem yields a set of coupled integral equations for the LFWFs.

\begin{figure}
\centering 
\raisebox{0.05\height}{\includegraphics[width=0.45\textwidth]{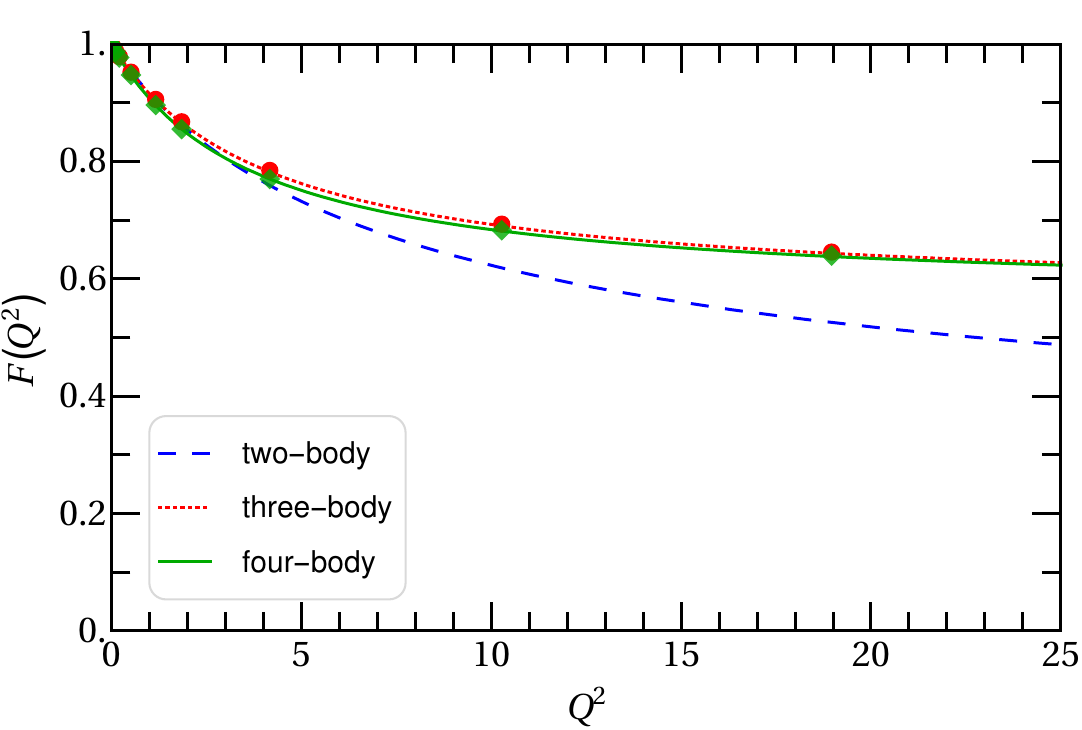}} \hfill
\includegraphics[width=0.47\textwidth]{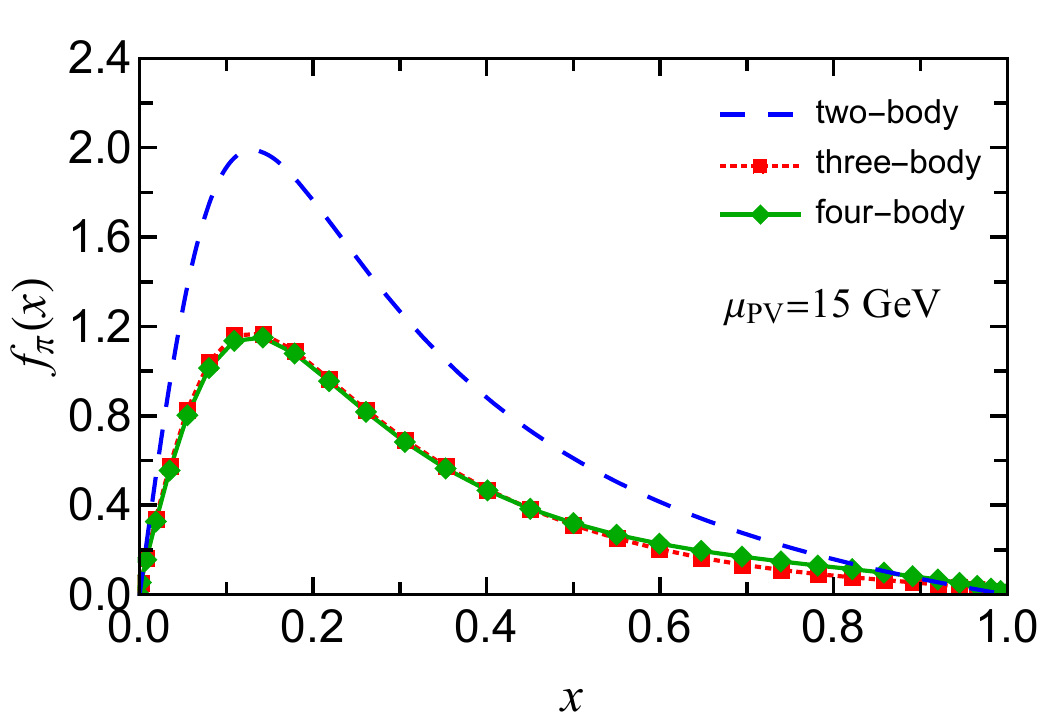}
\caption{Numerical convergence of the Fock sector expansion for two representative observables in the quenched scalar Yukawa theory at coupling $\alpha = 2.0$: (\textit{Left}) electromagnetic form factor $F(Q^2)$; (\textit{Right}) PDF  $f_\pi(x)$. The PV mass is set to be 15 GeV. The left panel uses units of $\mathrm{GeV}^2$
 on the horizontal axis; the PDF is evaluated at factorization scale $\mu_F = \infty$. The left panel is adapted from Ref.~\cite{Karmanov:2016yzu}, and the right panel is adapted from Ref.~\cite{Duan:2024dhy}. }
\label{fig:Fock_convergence_quenched}
\end{figure}

This model has been solved in the one-nucleon sector using Fock-space truncation up to three particles ($|N\rangle + |\pi N\rangle + |\pi\pi N\rangle$)~\cite{Karmanov:2008br, Karmanov:2016yzu} and four particles ($| N\rangle + |\pi N\rangle + |\pi\pi N\rangle + |\pi\pi\pi N\rangle$)~\cite{Li:2014kfa, Li:2015iaw}. As shown in Fig.~\ref{fig:Fock_convergence_quenched}, comparison between three- and four-body results \cite{Karmanov:2016yzu, Duan:2024dhy} demonstrates numerical convergence of the Fock expansion even in the strongly coupled regime ($\alpha = 1.0 \sim 2.0$). 
Note that these solutions do not include anti-particle d.o.f.~within the Fock space. This is referred to as quenched. 
Recently, an unquenched solution was obtained by including the three-body Fock sector $|N N\bar N\rangle$, i.e., $| N\rangle_\text{ph} = | N\rangle + | \pi N\rangle + | \pi\pi N\rangle + |N N\bar N\rangle$. In this case, the anti-nucleon component contributes only a negligible fraction of the total norm~\cite{Zhang:2025wli}. 

The hadronic state vector in the one-nucleon sector is expressed in momentum space as, 
\begin{equation}\label{eqn:hadronic_state_vector}
|\psi(p)\rangle = \sum_{\mathcal F} \int \big[\dd x_i \dd^2 k_{i\perp} \big]_{\mathcal F} \psi_{\mathcal F}(\{x_i, \vec k_{i\perp}\}) |\{p_i\}_{\mathcal F}\rangle,
\end{equation}
where, ${\mathcal F}$ labels a Fock sector, and 
\begin{equation}
[\dd x_i \dd^2 k_{i\perp}]_{\mathcal F} = \frac{1}{S_{\mathcal F}} \prod_{i} \frac{\dd x_i}{2x_i}\frac{\dd^2 k_{i\perp}}{(2\pi)^3} 2\delta(\sum_i x_i - 1) (2\pi)^3\delta^2(\sum_i \vec k_{i\perp}),
\end{equation}
is the invariant phase-space measure. The symmetry factor $S_{\mathcal F}=n!m!l!$ accounts for $n$ nucleons, $m$ anti-nucleons, and $l$ pions in sector $\mathcal F$. The LFWF $\psi_{\mathcal F}(\{x_i, \vec k_{i\perp}\})$  depends only on the longitudinal momentum fractions $x_i={p_i^+}/{p^+}$ and the relative transverse momenta $\vec k_{i\perp} = \vec p_{i\perp} - x_i \vec p_\perp$. The Fock state is constructed as, 
\begin{multline}
|\{p_i\}_{\mathcal F}\rangle = b^\dagger(p_1)b^\dagger(p_2)\cdots b^\dagger(p_n) d^\dagger(p_{n+1}) d^\dagger (p_{n+2}) \cdots \\
\times d^\dagger(p_{n+m}) a^\dagger(p_{n+m+1}) a^\dagger(p_{n+m+2}) \cdots a^\dagger(p_{n+m+l}) |0\rangle,
\end{multline}
 with all particles on-shell ($p_i^2 = m_i^2$). The state vector is normalized as 
 \begin{equation}
 \langle\psi(p')|\psi(p)\rangle=2p^+(2\pi)^3\delta^3(p-p'),
 \end{equation}
 which implies that the LFWFs satisfy the normalization condition, 
\begin{equation}
\sum_{{\mathcal F}}  
\int \big[\dd x_i \dd^2k_{i\perp}\big] \, \psi_{\mathcal F}(\{x_i, \vec k_{i\perp}\})\psi_{\mathcal F}^*(\{x_i, \vec k_{i\perp}\}) 
= \sum_{{\mathcal F}}  Z_{{\mathcal F}}  
= 1,
\end{equation}
where  $Z_{{\mathcal F}}$ is the probability of the system residing in Fock sector $\mathcal F$. 
For the physical nucleon, $Z_N = Z$ coincides with the field-strength renormalization constant. 

The transverse momentum dependent parton distributions (TMDs) provide a three-dimensional image of the hadron in momentum space \cite{Boussarie:2023izj}. 
The unpolarized TMD for the nucleon inside the nucleon is defined by the light-front correlator
\begin{equation}\label{eqn:TMD_operator_def}
f_{N}(x,\vec{k}_\perp) = \frac{1}{2}k^+ \int \dd y^- \dd^2 y_\perp\, e^{i k \cdot y} \,
\langle \psi(p) | N^\dagger(0) N(y) | \psi(p) \rangle \Big|_{k^+ = x p^+},
\end{equation}
with light-front coordinates ($y^+=0$) and phase factor $k \cdot y = \tfrac{1}{2}k^+ y^- - \vec{k}_\perp \cdot \vec{y}_\perp$ in symmetric frame $p_\perp = 0$.  
Analogous definitions apply to the anti-nucleon and pion:
\begin{align}
f_{\bar N}(x,\vec{k}_\perp) =\,&  \frac{1}{2}k^+ \int \dd y^- \dd^2 y_\perp\, e^{i k \cdot y} \,
\langle \psi(p) | N(0) N^\dagger(y) | \psi(p) \rangle \Big|_{k^+ = x p^+}, \label{eqn:TMD_antinucleon_operator} \\
f_{\pi}(x,\vec{k}_\perp) =\,& \frac{1}{2}k^+ \int \dd y^- \dd^2 y_\perp\, e^{i k \cdot y} \,
\langle \psi(p) | \pi(0) \pi(y) | \psi(p) \rangle \Big|_{k^+ = x p^+}. \label{eqn:TMD_pion_operator}
\end{align}

Inserting a complete set of Fock states between the field operators yields the LFWF representation of the nucleon TMD: 
\begin{equation}\label{eqn:TMD_nucleon}
f_{N}(x,\vec k_\perp) = \sum_{\mathcal F} \int \big[\dd x_i \dd^2 k_{i\perp}\big]_{\mathcal F} \sum_{j\in N} (2\pi)^3\delta(x-x_j)\delta^2(k_\perp - k_{j\perp}) \big| \psi_{\mathcal F}(\{x_i,\vec k_{i\perp}\}) \big|^2,
\end{equation}
where the inner sum runs over all nucleons within Fock sector $\mathcal F$, and the integration spans the phase space of the remaining partons. 
Similar expressions hold for the anti-nucleon and pion TMDs. These representations are manifestly positive-definite and reflect the probabilistic interpretation of $\big|\psi_{\mathcal F}\big|^2$. 
Integrating over transverse momentum recovers the collinear PDFs, 
\begin{equation}\label{eqn:PDF_operator_def}
\begin{split}
f_{N}(x) =\,& \frac{k^+}{4\pi } \int \dd y^- \, e^{\frac{i}{2} x p^+ y^-} \,
\langle \psi(p) | N^\dagger(0) N(y) | \psi(p) \rangle \\
=\,& \int \frac{\dd^2 k_\perp}{(2\pi)^3} f_{N}(x,\vec k_\perp),
\end{split}
\end{equation}
with analogous definitions for pion and anti-nucleon PDFs.

Unlike in QCD, no Wilson lines are required here: the scalar Yukawa model lacks gauge interactions, so the bilocal operators are automatically invariant. Consequently, the TMDs are free of gauge-link ambiguities. Nevertheless, as non-local operators, they retain dependence on the factorization scale and renormalization scheme. Thanks to super-renormalizability, UV divergences are confined to a finite set of diagrams, and the renormalization-group evolution is greatly simplified. In particular, because no UV divergences arise beyond one loop, we may safely take the factorization scale to infinity ($\mu_F \to \infty$), yielding scale-independent TMDs. The implications of a finite $\mu_F$ will be discussed in Sec.~\ref{sec:discussions}.

Although the LFWFs are probability amplitudes, the resulting TMDs and PDFs are not normalized probability distributions. For instance, 
\begin{align}
   &\int_0^1 \dd x \int \frac{\dd^2k_\perp}{(2\pi)^3} \big[ f_N(x, k_\perp) +  f_{\bar N}(x, k_\perp) +  f_\pi(x, k_\perp)\big] \\
= &\int_0^1 \dd x  \big[ f_N(x) +  f_{\bar N}(x) +  f_\pi(x)\big] \\
\ne & \,1.
\end{align}
Namely, they are not normalized to unity. 
Instead, they obey physical sum rules. Momentum conservation gives
\begin{equation}
\int_0^1 \dd x\, x \big[ f_N(x) +  f_{\bar N}(x) + f_\pi(x) \big] = 1.
\end{equation}
while baryon number conservation implies, 
\begin{equation}
\int_0^1 \dd x\, f^{v}_N(x) = 1.
\end{equation}
However, the valence distribution $f^{v}_N= f_N - f_{\bar N}$ is not guaranteed to be positive-definite, except in the quenched approximation where $f_{\bar N} = 0$. In the unquenched case, the physical nucleon is still  dominated by the bare nucleon and pion-nucleon fluctuations, and the anti-nucleon contributions remain numerically small, so  $f^{v}_N(x) \approx f_N(x)$.

To construct properly normalized distributions, one may define normalized TMDs as (cf. Refs.~\cite{Ehlers:2022oal, Ehlers:2022oke}), 
\begin{equation}
\hat f_i(x, k_\perp) \equiv \mathcal N^{-1}_i f_i(x, k_\perp)
\end{equation}
where, 
\begin{equation}
\mathcal N_i = \int \dd x \int \frac{\dd^2k_\perp}{(2\pi)^3} f_i(x, k_\perp).
\end{equation}
Integrating out the transverse momentum gives the normalized PDF $\hat f_i(x)$. 
On the other hand, the lack of unit normalization in the TMD $f_N(x, k_\perp)$ also stems from overcounting  identical particles. For example, in the $|NN\bar N\rangle$ sector, both nucleons contribute to $f_N$. To avoid this, we introduce the one-parton TMD (1PTMD):
\begin{equation}\label{eqn:OPTMD_nucleon}
\bar f_{N}(x,\vec k_\perp) = \sum_{\mathcal F} \int \big[\dd x_i \dd^2 k_{i\perp}\big]_{\mathcal F} (2\pi)^3\delta(x-x_N)\delta^2(k_\perp - k_{N\perp}) \Big| \psi_{\mathcal F}(\{x_i,\vec k_{i\perp}\}) \Big|^2,
\end{equation}
which counts only one nucleon per Fock sector and is guaranteed to be normalized. The corresponding one-parton PDF (1PPDF) is $\bar f_N(x) = \int \frac{\dd^2k_\perp}{(2\pi)^3} \bar f_N(x, k_\perp)$. 

\begin{figure}
\centering 
\includegraphics[width=0.47\textwidth]{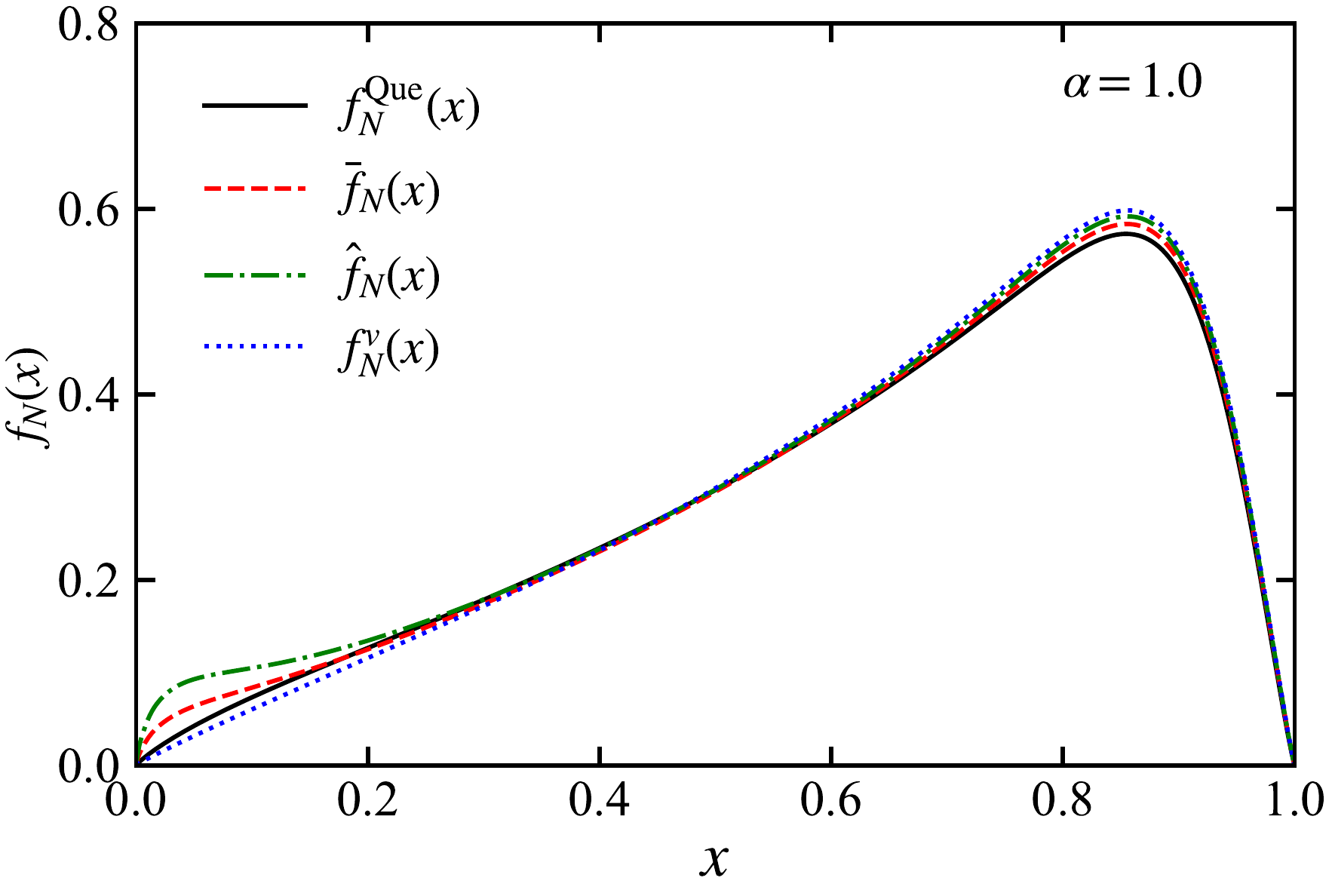} \hfill
\includegraphics[width=0.47\textwidth]{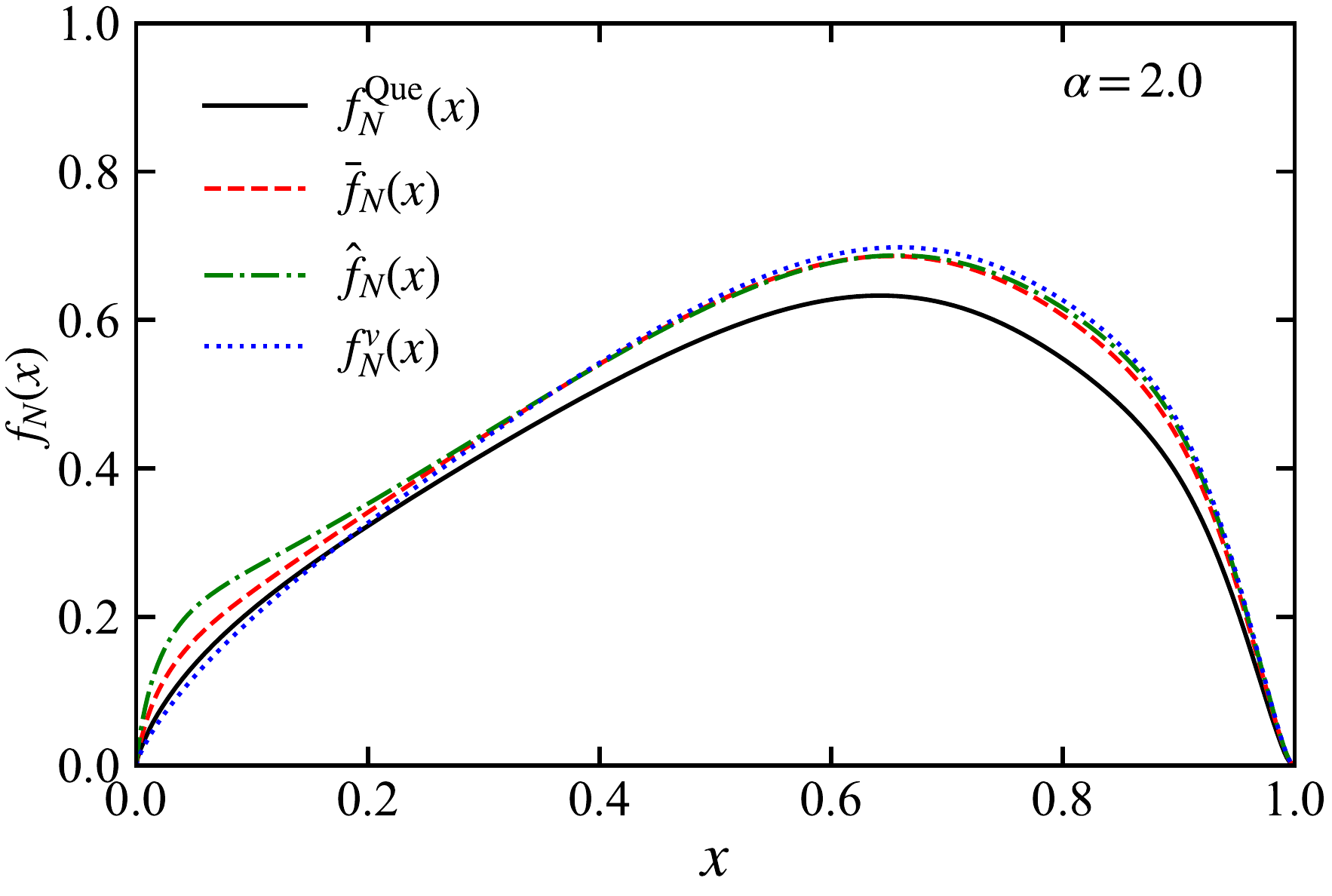}
\caption{Comparison of nucleon PDFs in the physical nucleon: quenched PDF $f^{\text{Que}}_N(x)$;  one-parton PDF $\bar f_N(x)$; normalized unquenched PDF $\hat f_N(x)$ and valence PDF $f^v_N(x)$. Results are based on a three-body Fock truncation (see texts) with a coupling (\textit{Left}): $\alpha = 1.0$; (\textit{Right}): $\alpha = 2.0$. The $\delta(1-x)$ contribution from the one-body sector is omitted for clarity. }
\label{fig:PDFs}
\end{figure}

In the quenched approximation, the valence, normalized, and one-parton PDFs all coincide with $f^{\text{Que}}_N(x)$. 
Fig.~\ref{fig:PDFs} compares these distributions in both quenched and unquenched calculations. 
At moderate coupling $\alpha = 1.0$, the differences are negligible; at strong coupling $\alpha = 2.0$, deviations become visible, reflecting the contribution of sea partons. 
As we will show later in Sec.~\ref{sec:entanglement_in_unquenched_theory} that the Shannon entropy of $\hat f(x, k_\perp)$ and $\bar f(x, k_\perp)$ are different from the parton entanglement entropy.

\section{Density matrices} \label{sec:density_matrix}

To quantify the quantum entanglement between the nucleon, pion, and anti-nucleon constituents, we partition the total Hilbert space into three subspaces labeled by particle species:
\begin{equation}
\mathcal{H} = \mathcal{H}_N \otimes \mathcal{H}_{\bar N} \otimes  \mathcal{H}_\pi,
\end{equation}
where $\mathcal{H}_N$ ($\mathcal{H}_{\bar N}$) is the Fock space containing states with only nucleons (anti-nucleons), and $\mathcal{H}_\pi$ is the Fock space of pions.

The reduced density matrix for the nucleon subsystem is obtained by tracing out the pion and anti-nucleon d.o.f.:
\begin{equation}
\rho_N = \operatorname{Tr}_{\pi,\bar N}  |\Psi\rangle\langle\Psi |. 
\end{equation}
The entanglement entropy quantifying the quantum correlations between the nucleon and the rest of the system is then given by the von Neumann entropy of $\rho_N$:
\begin{equation}
S_N  = S_\text{vN}(\rho_N) \equiv -\operatorname{Tr} \rho_N \log \rho_N.
\end{equation}
If $\rho_N$ is diagonal, i.e.,
\begin{equation}
\rho_N = \sum_\alpha p_\alpha |\alpha \rangle\langle \alpha |,
\end{equation}
the von Neumann entropy reduces to the Shannon entropy,
\begin{equation}
S_N  = H(\{p_\alpha\}) \equiv -\sum_\alpha p_\alpha \log p_\alpha.
\end{equation}
Analogous reduced density matrices $\rho_\pi$ and $\rho_{\bar N}$, and their corresponding entanglement entropies $S_\pi$ and $S_{\bar N}$, can be defined for the pion and anti-nucleon subsystems, respectively.

The density matrix of a pure state $\rho = |\Psi\rangle\langle\Psi|$ presupposes a normalized state vector, $\braket{\Psi | \Psi} = 1$. To construct such a state, we supplement the hadronic eigenstate Eq.~\eqref{eqn:hadronic_state_vector} with a wave packet:
\begin{equation}\label{eqn:normalized_state_vector}
    \ket{\Psi}=\int\frac{\dd^3p}{(2\pi)^32p^{+}}\Psi(p)\ket{\psi(p)},
\end{equation}
where the hadronic wave packet $\Psi(p)=\braket{p|\Psi}$ satisfies the normalization condition
\begin{equation}\label{eqn:wave_package}
    \int\frac{\dd^3p}{(2\pi)^32p^{+}}|\Psi(p)|^2 = 1.
\end{equation}
A typical choice is a Gaussian wave packet,
\begin{equation}\label{eqn:Gaussian_wavepacket}
    \Psi(p) = \mathcal N \sqrt{2p^{+}} \exp\!\Big[-\frac{(p-P_0)^2}{2\sigma^2}\Big],
\end{equation}
where $\mathcal N$ is a normalization constant. In the narrow wave packet limit $\sigma \to 0$, this becomes a sharply peaked Dirac-$\delta$:
\begin{equation}
|\Psi(p)|^2 \to 2p^+(2\pi)^3 \delta^{3}(p-P_0).
\end{equation}
The full state vector with the wave packet reads
\begin{equation}\label{eqn:state_vector_wavepacket}
|\Psi\rangle = \sum_{{\mathcal F}}\frac{1}{S_{\mathcal F}}\prod_{i}\int\frac{\dd^3p_i}{(2\pi)^32p^{+}_{i}} \Psi_{\mathcal F}(\{p_i\}) |\{p_i\}_{\mathcal F}\rangle.
\end{equation}
Here, the single-particle wave function $\Psi_{\mathcal F}(\{p_i\})$ factorizes into the intrinsic LFWFs and the hadronic wave packet:
\begin{equation}\label{eqn:LFWF with wave packet}
\Psi_{\mathcal F}(\{p_i\}) \equiv \Psi(p) \psi_{\mathcal F}(\{x_i, \vec k_{i\perp}\}),
\end{equation}
with total momentum $p = \sum_i p_i$, longitudinal momentum fractions $x_i = p^+_i/p^+$, and relative transverse momenta $\vec k_{i\perp} = \vec p_{i\perp} - x_i \vec p_\perp$.

The von Neumann entropy is conventionally defined for discrete spectra, yet our system involves continuous momentum variables $\{p_1, p_2, \ldots\}$. To adhere to the standard formalism, we discretize momenta by confining the system to a finite box:
\begin{equation}\label{eqn:box_regularization}
    -L \leq x^- \leq +L, \quad -\frac{L_\perp}{2} \le x^{1,2} \le +\frac{L_\perp}{2}.
\end{equation}
We follow the standard convention in discretized light-cone quantization (DLCQ) and adopt the periodic boundary conditions (PBC) in the transverse directions, while in
the longitudinal direction we apply PBC for bosons and anti-periodic boundary conditions
(ABC) for fermions \cite{Brodsky:1997de}. The allowed momenta become discrete:
\begin{align}
    p^{+} &= \frac{2\pi n}{L}, & & n = 0, 1, 2, \ldots \text{ (bosons)},\quad n = \tfrac{1}{2}, \tfrac{3}{2}, \ldots \text{ (fermions)}, \\
    \vec p_\perp &= \frac{2\pi}{L_\perp}(n_1, n_2), & & n_{1,2} = 0, \pm1, \pm2, \ldots
\end{align}
The momentum integration and Dirac $\delta$-function are then replaced by discrete sums:
\begin{equation}\label{eqn:discrete transformation}
    \int\frac{\dd^3p}{(2\pi)^32p^{+}} \to \frac{1}{V}\sum_{p^{+}}\sum_{\vec p_{\perp}}\frac{\theta(p^{+})}{2p^{+}}, \quad (2\pi)^3\delta^{3}(p-p') \to V\,\delta_{p,p'},
\end{equation}
where $V = L L_\perp^2$ is the spacetime volume and $\theta(z)$ is the Heaviside step function.
In the narrow wave packet limit $\sigma \to 0$, the Gaussian becomes
\begin{equation}\label{eqn:narrow wave packet limit}
\Psi(p) \to \sqrt{2p^+ V}\,\delta_{p,P_0}.
\end{equation}
The continuum limit is recovered as $L, L_\perp \to \infty$. For a hadron with definite longitudinal momentum $P_0^+$, we set $P_0^+ = (2\pi/L)K$, where $K$ is known as the light-cone harmonic resolution \cite{Pauli:1985ps, Pauli:1985pv}. Light-front boost invariance ensures that the intrinsic LFWFs depend only on the longitudinal momentum fractions $x_i = n_i/K$ and relative momentum $\vec k_{i\perp}$. In the continuum limit, $K \to \infty$ while $K/L$ remains finite. In what follows, we perform calculations in the discretized framework and take the continuum limit at the end.

With the normalized state vector~\eqref{eqn:state_vector_wavepacket}, the full density matrix reads
\begin{equation}\label{eqn:density_matrix}
    \rho = \sum_{{\mathcal F}, {\mathcal G}}\frac{1}{S_{\mathcal F}S_{\mathcal G}}\prod_{i\in {\mathcal F}} \int\frac{\dd^3p'_i}{(2\pi)^32p'^{+}_{i}} \prod_{j\in {\mathcal G}} \int\frac{\dd^3p_j}{(2\pi)^32p_{j}^{+}}
     \Psi_{\mathcal F}^*(\{p'_i\})\Psi_{\mathcal G}(\{p_j\}) |\{p'_i\}_{\mathcal F}\rangle\langle \{p_j\}_{\mathcal G} |.
\end{equation}
Here, $\mathcal F$ and $\mathcal G$ label Fock sectors. Tracing out pions and anti-nucleons yields the reduced density matrix for nucleons:
\begin{equation}\label{eqn:rho_N_unquenched}
    \rho_N  
    	  = \sum_{\mathcal N, \mathcal N'}\frac{1}{S_{\mathcal N}S_{\mathcal N'}}\prod_{i \in \mathcal N'} \int\frac{\dd^3p'_i}{(2\pi)^32p'^{+}_{i}} \prod_{j\in \mathcal N} \int \frac{\dd^3p_j}{(2\pi)^32p_{j}^{+}}
     \rho_{\mathcal N\mathcal N'}(\{p'_i\}, \{p_j\}) | \{p'_i\}_{\mathcal N'}\rangle\langle \{p_j\}_{\mathcal N} |,
\end{equation}
where $\mathcal N, \mathcal N'$ denote Fock sectors containing only nucleons, and $S_{\mathcal N} = n!$ for $n$ identical nucleons. The matrix elements are
\begin{multline*}
\rho_{\mathcal N\mathcal N'}(\{p'_i\}, \{p_j\}) = \sum_{\mathcal F}\frac{1}{S_\mathcal F} \prod_{i\in \mathcal F} \int\frac{\dd^3k_i}{(2\pi)^32k_{i}^{+}} \\
 \times  \Psi_{\mathcal N' \cup  \mathcal F}^*(\{k_1, k_2, \ldots, p'_1, p'_2, \ldots\}) 
\Psi_{\mathcal N \cup  \mathcal F}(\{k_1, k_2, \ldots, p_1, p_2, \ldots \}),
\end{multline*}
with the sum over $\mathcal F$ restricted to Fock sectors containing no nucleons (i.e., pions, anti-nucleons, or vacuum).
Note that $\rho_N$ is generally non-diagonal. Therefore, computing the entanglement entropy requires diagonalizing $\rho_N$ first.

\section{Entanglement in quenched scalar theory}\label{sec:entanglement_in_quenched_theory}

In this section, we consider the quenched approximation, in which nucleon-antinucleon pair creation is excluded, i.e., there are no sea nucleons or antinucleons. 
In this approximation, the reduced density matrix simplifies dramatically. The physical nucleon state can be schematically expanded as
\begin{equation}
    \ket{N}_\text{ph} = \ket{N} + \ket{\pi N} + \ket{\pi\pi N} + \cdots,
    \label{quenched expansion}
\end{equation}
where each term corresponds to a Fock sector with a fixed number of partons. 
As discussed in Sec.~\ref{sec:density_matrix}, the Fock expansion converges rapidly, with the three-body truncation ($|N\rangle + |\pi N\rangle + |\pi\pi N\rangle$) providing an accurate description even in the non-perturbative regime. Hence, we adopt this truncation for all practical calculations in this section.
Since the quenched theory contains exactly one nucleon in every Fock sector, it is convenient to label sectors by their total particle number: the $n$-body sector consists of one nucleon and $(n-1)$ pions.

It is then straightforward to show that the reduced density matrix for the nucleon subsystem takes the form
\begin{equation}\label{eqn:rho_N_quenched}
\rho_N^{\text{Que}} = \int\frac{\dd^3p_n}{(2\pi)^3 2p^{+}_{n}} \int\frac{\dd^3p'_n}{(2\pi)^3 2p'^{+}_{n}} \, \rho_N(p'_n, p_n) \, |p'_n\rangle\langle p_n|,
\end{equation}
with matrix elements
\begin{equation*}
\rho_N(p'_n, p_n) = \sum_{n=1}^\infty \frac{1}{(n-1)!} \prod_{i=1}^{n-1}\int\frac{\dd^3p_i}{(2\pi)^3 2p_{i}^{+}} \,
 \Psi_n^*(p_1, \dots, p_{n-1}, p'_n)\, \Psi_n(p_1, \dots, p_{n-1}, p_n).
\end{equation*}
Here, $\Psi_n$ denotes the LFWF for the $n$-body Fock sector, and the symmetry factor $1/(n-1)!$ accounts for the identical pions.
In the narrow Gaussian wave packet limit ($\sigma \to 0$), the hadronic wave function becomes sharply peaked: $|\Psi(p)|^2 \to 2p^+(2\pi)^3 \delta^{3}(p - P_0)$. Momentum conservation then enforces $p_n = p'_n$, rendering $\rho_N^{\text{Que}}$ diagonal in momentum space. The reduced density matrix simplifies to
\begin{equation}\label{eqn:rho_N_quenched_diagonal}
\rho_N^{\text{Que}} = \frac{1}{P^+_0 V} \int \frac{\dd x}{2x} \int \frac{\dd^2 k_\perp}{(2\pi)^3} \, f_N(x, {k}_\perp) \, |p\rangle\langle p|,
\end{equation}
where the on-shell nucleon momentum is parametrized as
\[
p^+ = x P^+_0, \quad \vec{p}_{\perp} = \vec{k}_\perp + x \vec{P}_{0\perp}, \quad p^2 = m^2,
\]
and the nucleon TMD is given by
\begin{equation}\label{eqn:TMD}
f_N(x, {k}_\perp) = \sum_{n} \int [\dd x_i \dd^2 k_{i\perp}]_n \, (2\pi)^3 \delta(x - x_n) \delta^2( {k}_\perp -  {k}_{n\perp}) \, \big|\psi_n(\{x_i, \vec{k}_{i\perp}\})\big|^2.
\end{equation}

Because sea partons are absent in the quenched approximation, there is no anti-nucleon contribution, and the nucleon TMD coincides with the valence distribution: $f_N^v = f_N$. Consequently, $f_N(x, {k}_\perp)$ is both positive definite and properly normalized:
\begin{equation}\label{eqn:TMD_normalization}
\int_0^1 \dd x \int \frac{\dd^2 k_\perp}{(2\pi)^3} \, f_N(x, {k}_\perp) = 1.
\end{equation}

From the diagonal form of the reduced density matrix~\eqref{eqn:rho_N_quenched_diagonal}, we can directly compute the von Neumann entropy, which is given by the Shannon entropy of the nucleon TMD:
\begin{align}\label{eqn:TMD_entropy}
    S_N = H(f_N)  
     =\,& -\int \dd x \int \frac{\dd^2k_\perp}{(2\pi)^3} f_N(x, {k}_{\perp})\log \Big[ \frac{1}{P_0^+V}f_N(x, {k}_{\perp}) \Big], \\
    \label{eqn:TMD_entropy2}
    =\,&\log (P^+_0V) - \int \dd x \int \frac{\dd^2k_\perp}{(2\pi)^3} f_N(x, {k}_{\perp})\log  f_N(x, {k}_{\perp}). 
\end{align}
The above expression contains a logarithmic IR divergence $\log P^+_0V$, reflecting the infinite number of d.o.f. in the continuum limit. 
It is instructive to restore the discrete version of the Shannon entropy:
\begin{equation}\label{eqn:TMD_entropy_discrete}
    S_N 
     =  \log (P_0^+V) - \frac{1}{P_0^+V}\sum_{x, \vec{k}_\perp} f_N(x,  {k}_{\perp})\log f_N(x,  {k}_{\perp}).
\end{equation}
For the scalar Yukawa model, the second term in Eq.~\eqref{eqn:TMD_entropy_discrete} converges in the continuum limit $V\to \infty$, except for the singular one-body contribution. Hence, it is practical to work directly with the continuous form~\eqref{eqn:TMD_entropy}.

For the quenched scalar Yukawa theory, it is useful to decompose the TMD according to Fock sectors, analogous to the LFWF expansion. We denote by $f_N^{(n)}$ the contribution from the $n$-body sector (one nucleon plus $n-1$ pions). The one-body term represents the point-like core of the physical nucleon:
\begin{equation}\label{eqn:TMD_f1}
    f_N^{(1)}(x,  {k}_{\perp}) = Z(2\pi)^3 \delta(x-1) \delta^2({k}_{\perp}).
\end{equation}
Note that this term introduces an additional IR divergence, since $(2\pi)^3 \delta(x-1) \delta^2( {k}_{\perp}) \to P_0^+V\, \delta_{p, P_0}$ in the discretized theory.

Starting from the two-body sector, the TMD encodes the pion cloud surrounding the nucleon. The general $n$-body contribution reads
\begin{multline}\label{eqn:TMD_fn}
    f_N^{(n)}(x_n,  {k}_{n\perp})=\frac{1}{(n-1)!}
    \prod_{i=1}^{n-1} \int \frac{\dd x_i}{2x_i} \int\frac{\dd^2k_{i\perp}}{(2\pi)^3} \\
    \times 2\delta\Big(\sum_{i=1}^n x_i - 1\Big) (2\pi)^3\delta^2\Big(\sum_{i=1}^n \vec{k}_{i\perp}\Big) 
\big| \psi_{n}(\{x_i, \vec{k}_{i\perp}\}) \big|^2.
\end{multline}
For example, the two-body component simplifies to
\begin{equation}\label{eqn:TMD_f2}
    f_N^{(2)}(x,  {k}_{\perp})= \frac{|\psi_{\pi N}(1-x,  {k}_{\perp})|^2}{2x(1-x)}.
\end{equation}
Substituting (\ref{eqn:TMD_fn}) into (\ref{eqn:TMD_entropy2}), we obtain
\begin{multline}  \label{eqn:EE_for_sykw}   
      S_N=\log(P_0^+V) - \int_0^1\text{d}x\int\frac{\text{d}^2k_{\perp}}{(2\pi)^3} \Big[f_N^{(1)}(x,k_{\perp})
      +\sum_{n\geq2}f_N^{(n)}(x,k_{\perp})\Big] \\
      \times \log\Big[f_N^{(1)}(x,k_{\perp})+\sum_{n\geq2}f_N^{(n)}(x,k_{\perp})\Big].
  \end{multline}
Taking into account the fact that $f_N^{(1)}(x,  {k}_{\perp}) = Z(2\pi)^3 \delta(x-1) \delta^2({k}_{\perp})$ and $f_N^{(n)}(x \to 1,  {k}_\perp) \to 0$ for $n\ge 2$, the $n=1$ term can be analytically evaluated and the entanglement entropy becomes,
  \begin{multline}    
   S_N = (1-Z)\log(P_0^+V)-Z \log Z \\ 
      - \int_0^1\dd x \int\frac{\dd^2k_{\perp}}{(2\pi)^3}\sum_{n \ge 2}f_N^{(n)}(x,  {k}_{\perp}) \log \Big[\sum_{n \ge 2}f_N^{(n)}(x,  {k}_{\perp})\Big].
  \end{multline}
Here, the $f_N^{(1)}(x,k_{\perp})$ in the logarithm in (\ref{eqn:EE_for_sykw}) needs to use the discretized version $f_N^{(1)}(x,k_{\perp})$ $=ZP_0^+V\delta_{p,P_0}$. The corresponding discrete form for finite volume $V$ is
\begin{multline}\label{eqn:EE_for_sykw_discrete}
    S_N=(1-Z)\log(P_0^+V)-Z \log Z \\ 
    - \frac{1}{P^+_0V}\sum_{x, \vec{k}_\perp} 
    \sum_{n \ge 2}f_N^{(n)}(x,  {k}_{\perp}) \log \Big[\sum_{n \ge 2}f_N^{(n)}(x,  {k}_{\perp})\Big].
\end{multline}
As expected, in the free theory ($\alpha = 0$), we have $Z = 1$ and the entanglement entropy vanishes.

\begin{figure}
\centering 
\includegraphics[width=0.47\textwidth]{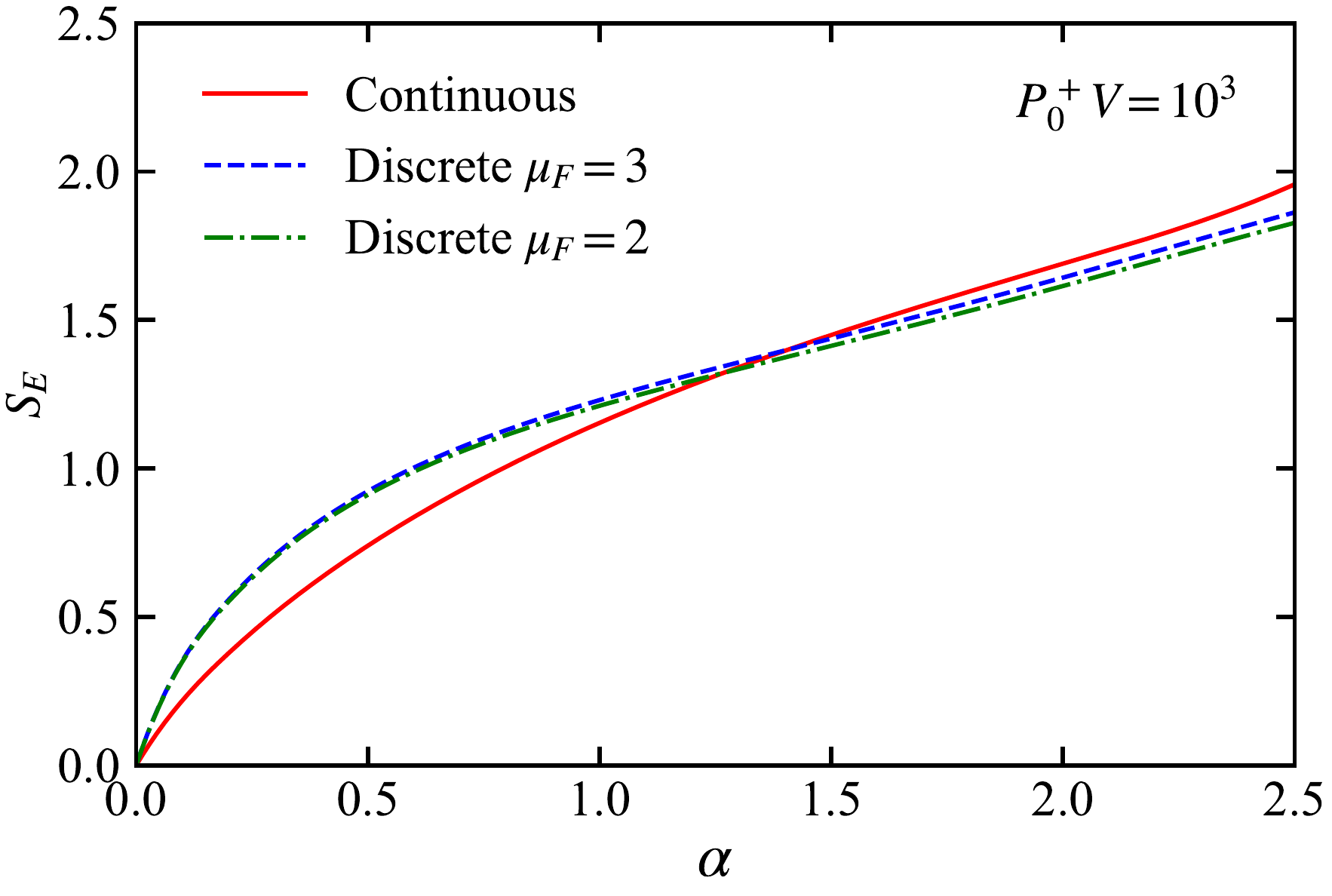}\hfill
\includegraphics[width=0.46\textwidth]{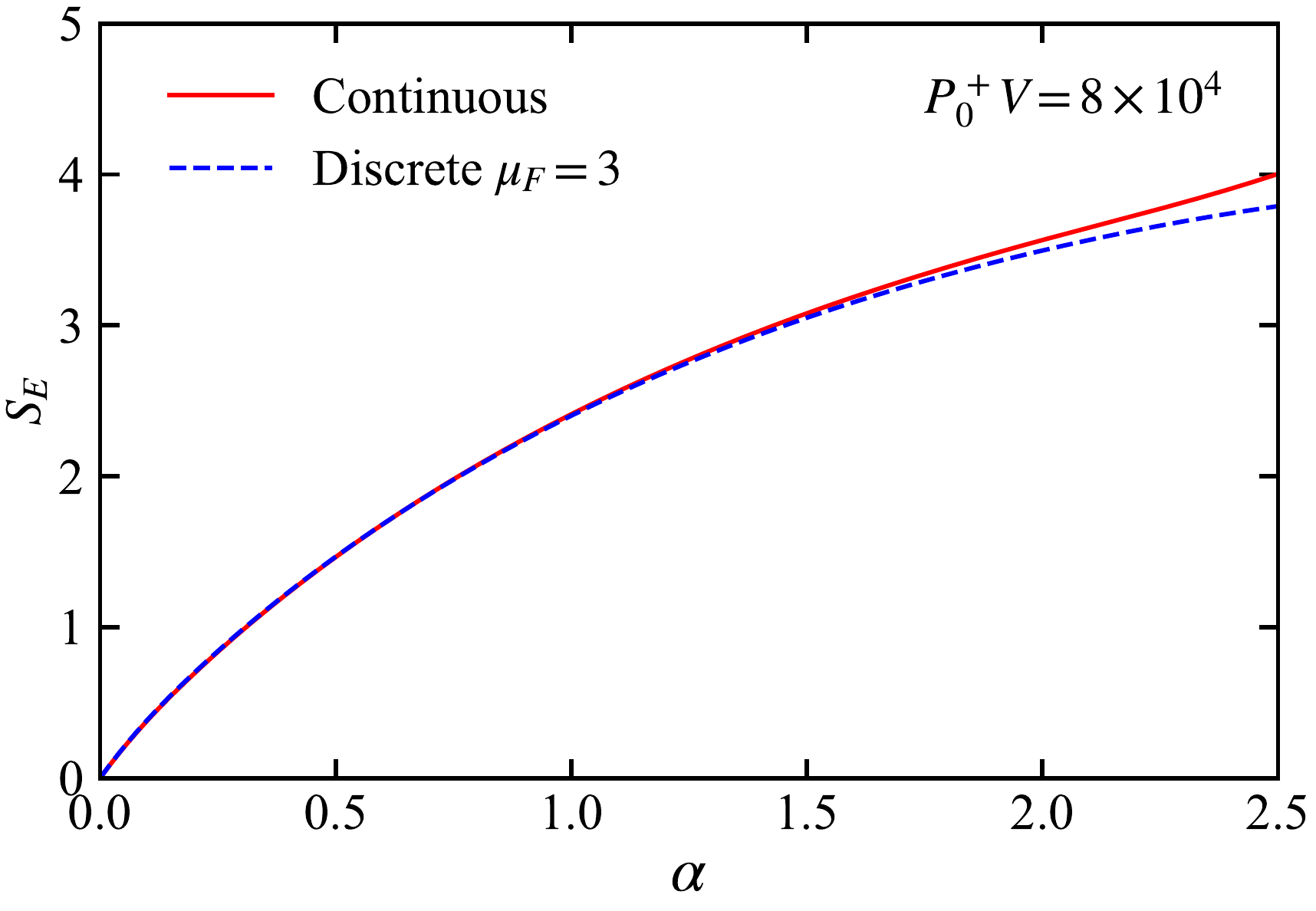}
\caption{The entanglement entropy $S_E \equiv S_N = S_\pi$ in the quenched theory as a function of the coupling $\alpha$. Since $S_E$ contains a logarithmic divergence $\log P^+_0V$, we fix $P^+_0V = 10^3$ and $8\times 10^4 \,\mathrm{GeV}^{-2}$. The ``continuous'' result uses Eq.~\eqref{eqn:EE_for_sykw}; the ``discrete'' result uses Eq.~\eqref{eqn:EE_for_sykw_discrete} with box regularization (see Sec.~\ref{sec:density_matrix}) and $P^+_0 = 1\,\mathrm{GeV}$. A transverse UV cutoff $\mu_F$ is applied in the discrete sums, as labeled. All quantities are in GeV units.}
\label{fig:quenched_SvN_discrete}
\end{figure}

Figure~\ref{fig:quenched_SvN_discrete} shows the entanglement entropy $S_E \equiv S_N = S_\pi$ as a function of the coupling $\alpha$ in the quenched theory with three-body truncation. The IR regulator is fixed to $P_0^+V = 10^3\,\mathrm{GeV}^{-2}$ (left panel) and $8\times 10^4\,\mathrm{GeV}^{-2}$ (right panel). The entropy increases monotonically with $\alpha$, as expected from stronger pion dressing. 
We compare two computational approaches: the ``continuous'' method uses Eq.~\eqref{eqn:EE_for_sykw} with continuum integrals, while the ``discrete'' method employs Eq.~\eqref{eqn:EE_for_sykw_discrete} with sums over a finite momentum grid, using $P_0^+ = 1\,\mathrm{GeV}$ and a transverse UV cutoff $\mu_F$ (indicated in the legend). Due to the simplicity of the scalar theory, the entropy converges rapidly with respect to $\mu_F$, and the finite part of Eq.~\eqref{eqn:EE_for_sykw_discrete} also stabilizes quickly as $V$ increases. Hence, we adopt the continuous formulation for subsequent calculations.

\begin{figure}
\centering 
\includegraphics[width=0.47\textwidth]{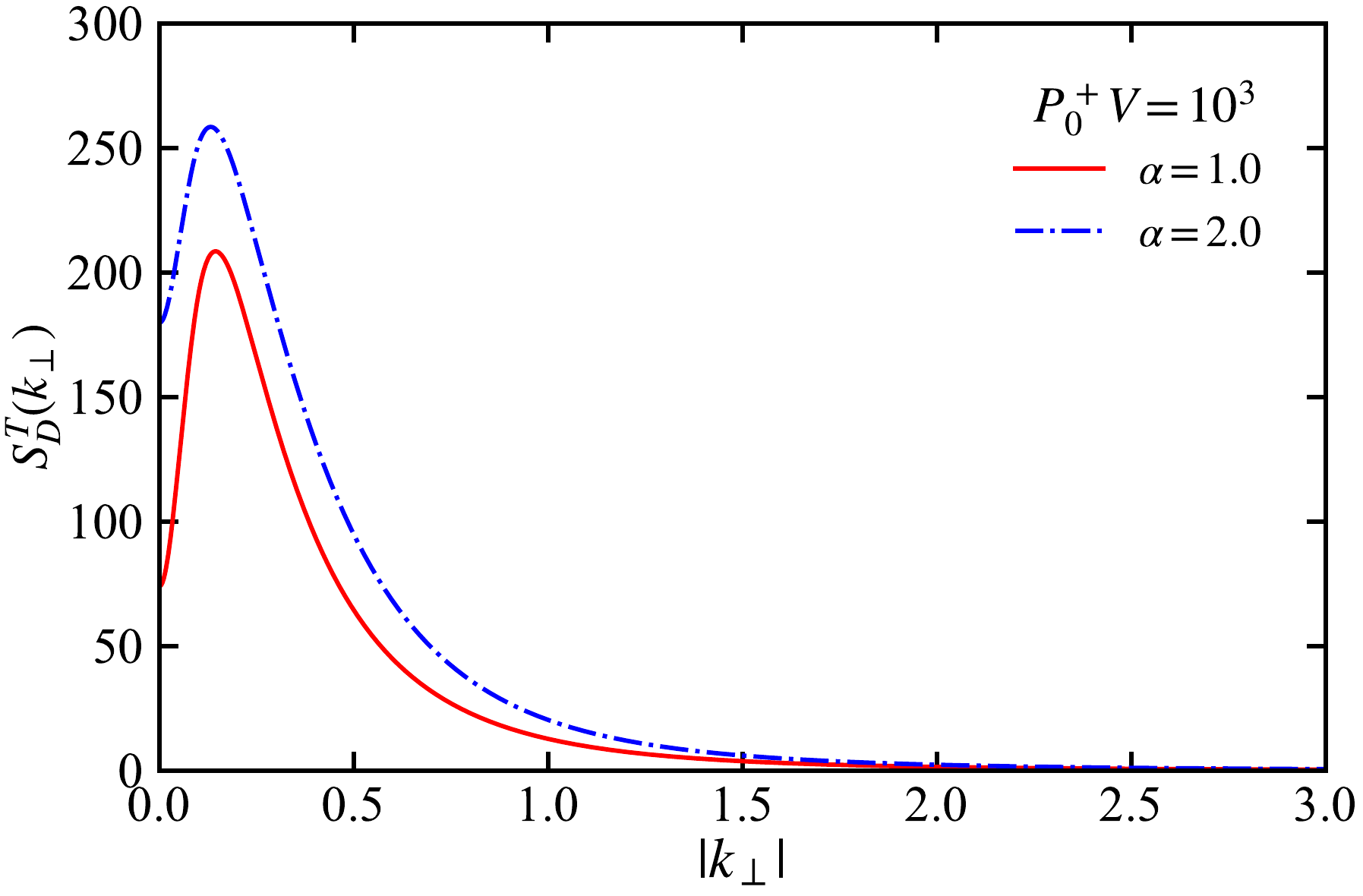}\hfill
\includegraphics[width=0.465\textwidth]{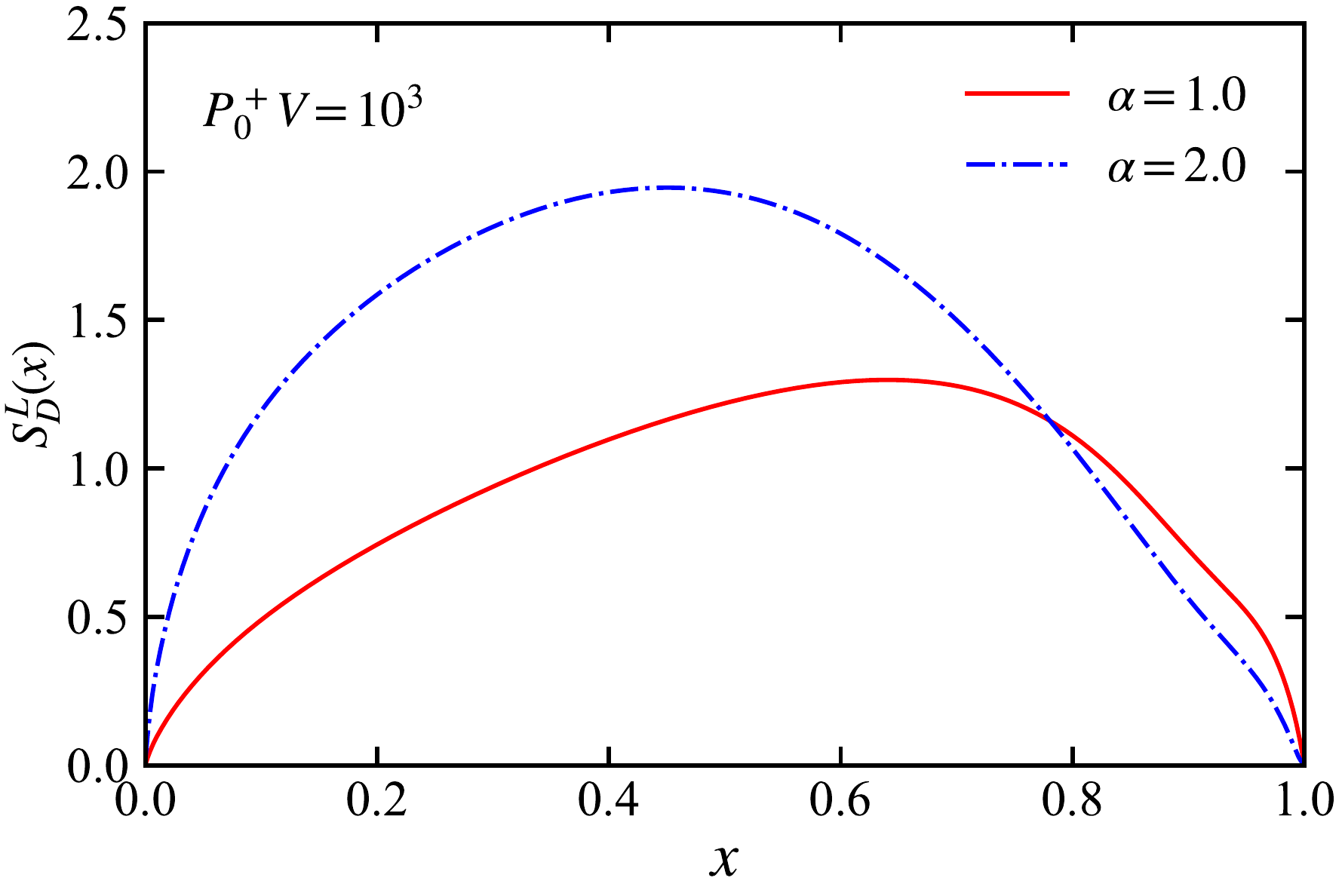}
\caption{Distribution of entanglement entropy in transverse and longitudinal momentum for $\alpha = 1.0$ and $2.0$, based on quenched three-body truncation with $P^+_0V = 10^3\,\mathrm{GeV}^{-2}$. The transverse peak near $0.1\,\mathrm{GeV}$, close to the pion mass scale. Note that radial phase-space factor $k_\perp$ is not included in the plot. The $\delta(1-x)$ one-body contribution is omitted for visibility.}
\label{fig:quenched_SvN_density}
\end{figure}

To understand how entanglement is distributed across momentum space, we define the transverse entropy density as
\begin{equation}
S_D^T({k}_\perp) = - \int \dd x\, f_N(x,  {k}_\perp) \log \Big[\frac{1}{P_0^+V}f_N(x,  {k}_\perp)\Big],
\end{equation}
and the longitudinal entropy density as (see, cf. \cite{Hagiwara:2017uaz, Bloss:2025ywh})
\begin{equation}
S_D^L(x) = -\int \frac{\dd^2k_\perp}{(2\pi)^3} f_N(x,  {k}_\perp) \log \Big[\frac{1}{P_0^+V}f_N(x,  {k}_\perp)\Big].
\end{equation}
Integration of either density over its argument reproduces the total entanglement entropy.
Note that $S_D^L(x)$ differs from the Shannon entropy density of the collinear PDF:
\begin{align}
H_D(x) =\,& - f_N(x) \log \Big(\frac{1}{2\pi K} f_N(x)\Big) \\
=\,& -\int \frac{\dd^2k_\perp}{(2\pi)^3} f_N(x,  {k}_\perp) \log \Big[\frac{1}{P^+_0L}\int \frac{\dd^2k_\perp}{(2\pi)^3} f_N(x,  {k}_\perp) \Big].
\end{align}
Fig.~\ref{fig:quenched_SvN_density} shows these densities for $\alpha = 1.0$ and $2.0$. The transverse entropy peaks near ${k}_\perp \approx 0.1\,\mathrm{GeV}$, close to the pion mass, while the longitudinal peak shifts with coupling, mirroring the evolution of the PDF shape.

Beyond entanglement entropy, other witnesses provide complementary insights. In the quenched theory, the system is bipartite (nucleon + pion), so the mutual information is simply $I(N:\pi) = 2S_N$. The linear entropy,
\begin{equation}
    S_L(\rho_A) = 1 - \operatorname{Tr}(\rho_A^2),
\end{equation}
quantifies the purity of subsystem $A$: it vanishes for pure states and increases with entanglement. For a pure global state, any mixedness in $\rho_A$ arises exclusively from entanglement with the complement.

\begin{figure}
    \centering
    \includegraphics[width=0.5\textwidth]{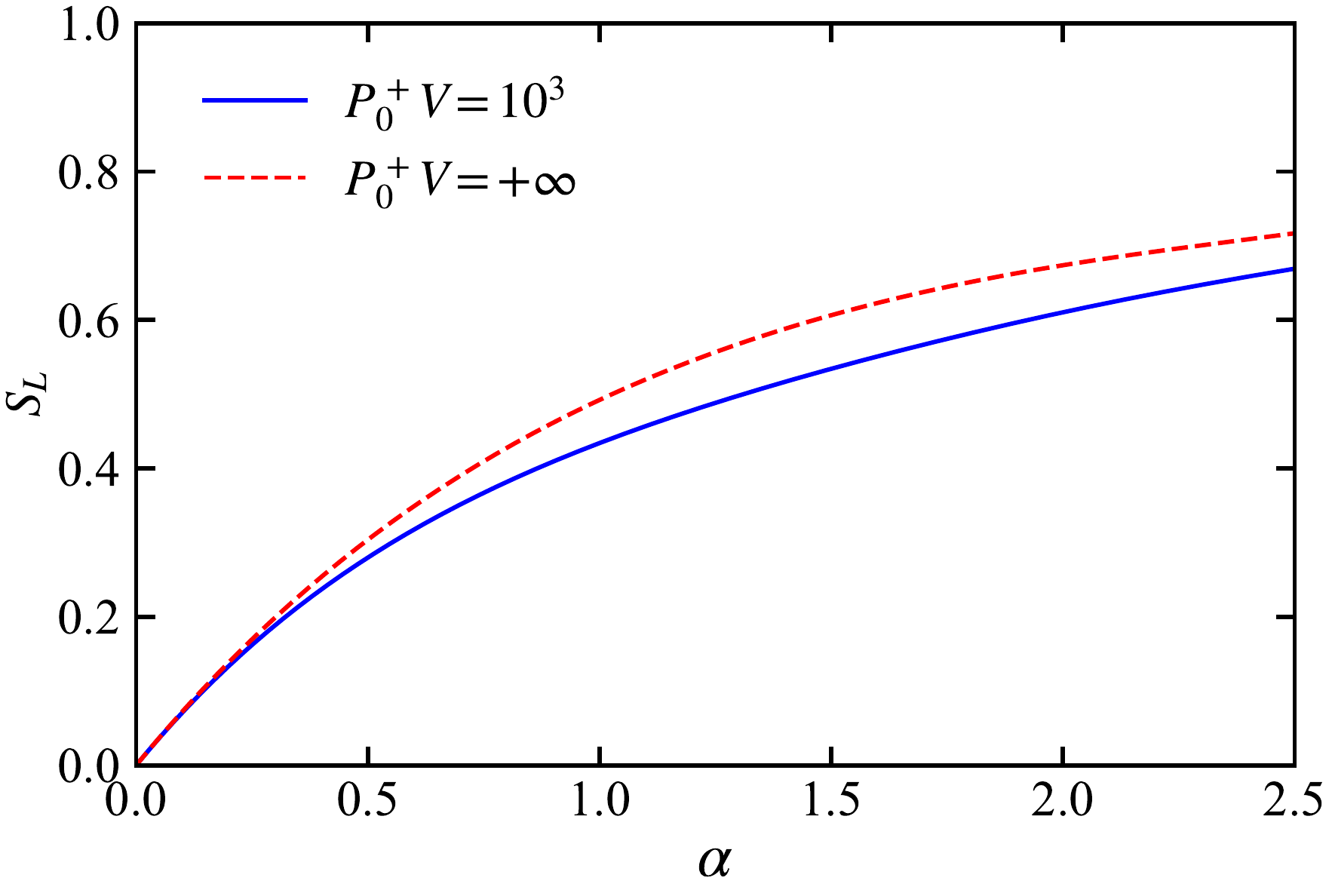}
    \caption{Linear entropy as a function of $\alpha$ in the quenched three-body theory. In the continuum limit, $S_L \to 1 - Z^2$, providing a finite entanglement measure.}
    \label{fig:linear_entropy_quenched}
\end{figure}

In the narrow wave packet limit, $\rho_N$ is diagonal, and the linear entropy reduces to
\begin{equation}
    S_L = 1-Z^2-\frac{1}{P_0^+V}\int \dd x \int\frac{\dd^2k_{\perp}}{(2\pi)^3}\Bigg[\sum_{n \geq 2}f_N^{(n)}(x,  {k}_{\perp})\Bigg]^2.
\label{linear entropy}
\end{equation}
Crucially, $S_L$ remains finite in the continuum limit: $\lim_{V\rightarrow\infty}S_L = 1 - Z^2$. This behavior is illustrated in Fig.~\ref{fig:linear_entropy_quenched}.

\section{Entanglement in unquenched scalar theory} \label{sec:entanglement_in_unquenched_theory}

In the previous section, we computed the entanglement between the nucleon and the pion cloud in the quenched theory, where nucleon-antinucleon pair creation is excluded. In that setting, the reduced density matrix $\rho_N$ is diagonal in the narrow wave packet limit ($\sigma \to 0$), and the entanglement entropy coincides with the Shannon entropy of the nucleon TMD. Here, we extend the analysis to the unquenched theory, which includes anti-nucleon d.o.f. Specifically, we consider the physical nucleon state truncated at three partons:
\begin{equation}\label{eq:Fock_expansion_unquenched}
    |{N}\rangle_\text{ph} = |{N}\rangle + |{\pi N}\rangle + |{\pi\pi N} \rangle + |{N N \bar{N}}\rangle.
\end{equation}

We then construct the full density matrix
\begin{equation}
    \rho = |{N}\rangle_\text{ph}\langle{N}|_\text{ph},
\end{equation}
assuming the Gaussian wave packet and box regularization introduced in Sec.~\ref{sec:density_matrix}.
The reduced density matrix for the nucleon subsystem is obtained by tracing out pions and anti-nucleons:
\begin{multline}\label{eq:rho_N_unquenched}
    \rho_{N} = \operatorname{Tr}_{\pi\bar{N}} \rho 
    = |N\rangle\langle N| + \operatorname{Tr}_{\pi} |\pi N\rangle\langle \pi N| + \operatorname{Tr}_{\pi} |\pi \pi N\rangle\langle \pi \pi N| \\
    + \operatorname{Tr}_{\bar{N}} |N N \bar{N}\rangle\langle N N \bar{N}|.
\end{multline}
Cross terms such as $\operatorname{Tr}_{\pi\bar N}|\pi N\rangle\langle N N \bar N|$ vanish due to the orthogonality between Fock sectors with different particle content.\footnote{While all non-Fock-diagonal terms vanish in the three-body truncation, they may persist in more general theories.}
In the narrow wave packet limit ($\sigma \to 0$), the first three terms become diagonal due to momentum conservation, as in the quenched case. However, the last term remains non-diagonal because two nucleons survive the trace over $\bar N$.

Crucially, the sector $\operatorname{Tr}_{\bar{N}} |N N \bar{N}\rangle\langle N N \bar{N}|$ is orthogonal to the others, as it contains two nucleons rather than one. Consequently, $\rho_N$ decomposes into a {direct sum} of two blocks:
\begin{equation}\label{eq:rho_N_unquenched_2}
\begin{split}
    \rho_{N} 
    =\,& \big(|N\rangle\langle N| + \operatorname{Tr}_{\pi} |\pi N\rangle\langle \pi N| + \operatorname{Tr}_{\pi} |\pi \pi N\rangle\langle \pi \pi N| \big)
    \oplus \operatorname{Tr}_{\bar{N}} |N N \bar{N}\rangle\langle N N \bar{N}| \\
    \equiv\,& \rho_N^{(1)} \oplus \rho_N^{(2)}.
\end{split}
\end{equation}
This block-diagonal structure allows us to diagonalize $\rho_N^{(1)}$ and $\rho_N^{(2)}$ independently. The total entanglement entropy thus splits additively:
\begin{equation}
    S_N \equiv S_\text{vN}(\rho_N) = -\operatorname{Tr}(\rho_N^{(1)}\log\rho_N^{(1)})-\operatorname{Tr}(\rho_N^{(2)}\log\rho_N^{(2)}).
\end{equation}
As noted, $\rho_N^{(1)}$ is already diagonal in the narrow wave packet limit. We now focus on $\rho_N^{(2)} = \operatorname{Tr}_{\bar{N}} |N N \bar{N}\rangle\langle N N \bar{N}| \equiv \operatorname{Tr}_{\bar{N}} \varrho$, where $\varrho \equiv |N N \bar{N}\rangle\langle N N \bar{N}|$.

The operator $\varrho$ acts on a Hilbert space containing two nucleons and one anti-nucleon. For such a pure bipartite state, the entanglement entropies of the reduced states satisfy
\begin{equation}
\operatorname{Tr}(\varrho_N\log\varrho_N) = \operatorname{Tr}(\varrho_{\bar N}\log\varrho_{\bar N}),
\end{equation}
where $\varrho_N = \operatorname{Tr}_{\bar{N}} \varrho = \rho_N^{(2)}$ and $\varrho_{\bar N} = \operatorname{Tr}_{N} \varrho$. Although $\varrho_N$ is non-diagonal, $\varrho_{\bar N}$ becomes diagonal in the narrow wave packet limit because only a single anti-nucleon remains after tracing out both nucleons:
\begin{equation}\label{eqn:varrho_Nbar_diagonal}
    \varrho_{\bar N} = \frac{1}{P^+_0 V} \int \frac{\dd x}{2x} \int \frac{\dd^2 k_\perp}{(2\pi)^3} f_{\bar N}(x,  {k}_\perp) \, |p\rangle\langle p|,
\end{equation}
with on-shell momentum $p^+ = x P^+_0$, $\vec{p}_{\perp} = \vec{k}_\perp + x \vec{P}_{0\perp}$, and $p^2 = m^2$. The anti-nucleon TMD is given by
\begin{equation}\label{eqn:Nbar_TMD}
    f_{\bar N}(x,  {k}_\perp) = \frac{1}{2!} \int_0^{1-x} \frac{\dd y}{4 x y (1 - x - y)} \int \frac{\dd^2 l_{\perp}}{(2\pi)^3} \big| \psi_{N N \bar{N}}(y, \vec{l}_{\perp}, 1 - x - y, -\vec{k}_{\perp} - \vec{l}_{\perp}) \big|^2.
\end{equation}
The corresponding entanglement entropy is therefore
\begin{equation}
\operatorname{Tr}(\varrho_{\bar N}\log\varrho_{\bar N}) = -Z_{N N \bar N} \log (P_0^+ V) + \int \dd x \int \frac{\dd^2 k_\perp}{(2\pi)^3} f_{\bar N}(x,  {k}_\perp) \log f_{\bar N}(x,  {k}_\perp).
\end{equation}
Combining both blocks, the total nucleon entanglement entropy reads
\begin{multline}\label{eqn:SE_rhoN_unquenched}
    S_N = (1 - Z) \log (P_0^+ V) - Z \log Z \\
    - \int \dd x \int \frac{\dd^2 k_{\perp}}{(2\pi)^3} \Big\{ f_N^{(\pi^n N)}(x,  {k}_{\perp}) \log f_N^{(\pi^n N)}(x,  {k}_{\perp}) + f_{\bar N}(x,  {k}_\perp) \log f_{\bar N}(x,  {k}_{\perp}) \Big\},
\end{multline}
where the multi-pion contribution is
\begin{multline}\label{eq:multi-pi_TMD}
    f_N^{(\pi^n N)}(x,  {k}_\perp) \equiv f_{N}^{(\pi N)}(x,  {k}_{\perp}) + f_{N}^{(\pi \pi N)}(x,  {k}_{\perp}) \\
    = \frac{\big|\psi_{\pi N}(1 - x, -\vec{k}_{\perp})\big|^2}{2 x (1 - x)}  
     + \frac{1}{2!} \int_0^{1 - x} \frac{\dd y}{4 x y (1 - x - y)} \int \frac{\dd^2 l_{\perp}}{(2\pi)^3} \\
     \times \big|\psi_{\pi\pi N}(y, \vec{l}_{\perp}, 1 - x - y, -\vec{k}_{\perp} - \vec{l}_{\perp})\big|^2.
\end{multline}

It is instructive to compare this entanglement entropy with the Shannon entropy $H(f_N)$ of the full nucleon TMD. Using the LFWF representation~\eqref{eqn:TMD_nucleon}, the TMD in the unquenched three-body truncation is
\begin{equation}
\begin{split}
    f_N(x,  {k}_\perp) =\,& f_N^{(1)}(x,  {k}_\perp) + f_N^{(\pi N)}(x,  {k}_\perp) + f_N^{(\pi \pi N)}(x,  {k}_\perp) + f_N^{(N N \bar N)}(x,  {k}_\perp) \\
    =\,& Z (2\pi)^3 \delta(x - 1) \delta^2( {k}_\perp) + \frac{\big|\psi_{\pi N}(1 - x, {k}_{\perp})\big|^2}{2 x (1 - x)} \\
      &+ \frac{1}{2!} \int_0^{1 - x} \frac{\dd y}{4 x y (1 - x - y)} \int \frac{\dd^2 l_{\perp}}{(2\pi)^3} \big|\psi_{\pi\pi N}(y, \vec{l}_{\perp}, 1 - x - y, -\vec{k}_{\perp} - \vec{l}_{\perp})\big|^2 \\
      &+ \int_0^{1 - x} \frac{\dd y}{4 x y (1 - x - y)} \int \frac{\dd^2 l_{\perp}}{(2\pi)^3} \big| \psi_{N N \bar{N}}(x, \vec{k}_{\perp}, y, \vec{l}_{\perp}) \big|^2.
\end{split}
\end{equation}
In the quenched limit, the sea contributions $f^{(N N \bar N)}_N$ and $f_{\bar N}$ vanish, and $S_\text{vN}(\rho_N)$ reduces precisely to $H(f_N)$. However, in the unquenched theory, $f_N(x,  {k}_\perp)$ is {not} normalized to unity (see Sec.~\ref{sec:lf_wf}), so $H(f_N)$ is ill-defined. Although one may construct normalized variants, such as the valence, one-parton, or normalized TMDs, the entanglement entropy~\eqref{eqn:SE_rhoN_unquenched} does not correspond to the Shannon entropy of any of these. This underscores that quantum entanglement encodes information beyond the classical information carried by the parton distributions \cite{Kharzeev:2021nzh}.

Similarly, the anti-nucleon entanglement entropy $S_{\bar N} = S_\text{vN}(\rho_{\bar N})$ follows from
\begin{equation}
\begin{split}
    \rho_{\bar N} =\,& \operatorname{Tr}_{N \pi} \rho \\
     =\,& 
\operatorname{Tr}_{N\pi} \big(|N\rangle + |\pi N\rangle + |\pi\pi N\rangle \big)\big(\langle N | + \langle \pi N | + \langle \pi\pi N | \big)
+ \operatorname{Tr}_N |NN\bar N\rangle\langle NN\bar N| \\
    =\,& (Z_N + Z_{\pi N} + Z_{\pi \pi N}) |0\rangle\langle 0| \oplus \varrho_{\bar N},
\end{split}
\end{equation}
where $|0\rangle$ is the Fock vacuum. The resulting entropy is
\begin{multline}\label{eqn:SE_rhoNbar_unquenched}
    S_\text{vN}(\rho_{\bar{N}}) = Z_{N N \bar{N}} \log (P_0^+ V) - (1 - Z_{N N \bar{N}}) \log (1 - Z_{N N \bar{N}}) \\
    - \int \dd x \int \frac{\dd^2 k_{\perp}}{(2\pi)^3} f_{\bar N}(x,  {k}_{\perp}) \log f_{\bar N}(x,  {k}_{\perp}).
\end{multline}

\begin{figure}
\centering
\includegraphics[width=0.47\textwidth]{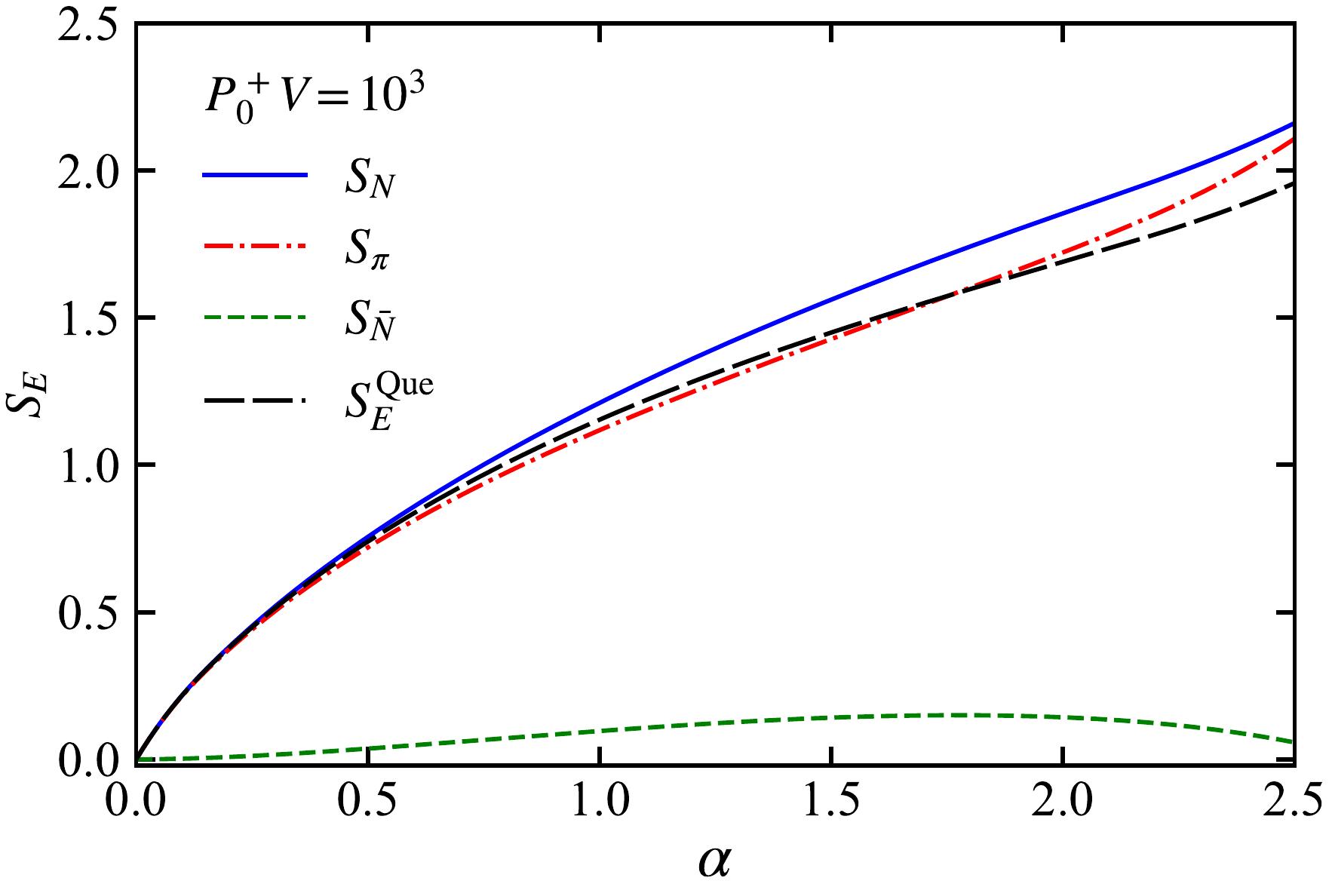}\hfill
\caption{Entanglement entropies $S_N$, $S_\pi$, and $S_{\bar N}$ in the unquenched three-body theory as functions of $\alpha$. The quenched result $S_E^{\text{Que}}$ is shown for reference. Here, $P^+_0 V = 10^3\,\mathrm{GeV}^{-2}$. }
\label{fig:EE_unquenched}
\end{figure}

For the pion subsystem, we note that $S_\text{vN}(\rho_\pi) = S_\text{vN}(\rho_{N \bar N})$, where $\rho_{N \bar N} = \operatorname{Tr}_\pi \rho$. A straightforward calculation yields
\begin{equation}
    \rho_{N \bar N} = |N\rangle\langle N| + \operatorname{Tr}_\pi (|\pi N\rangle\langle \pi N| + |\pi \pi N\rangle\langle \pi \pi N|) + |N N \bar N\rangle\langle N N \bar N|.
\end{equation}
Its entropy is
\begin{multline}\label{eqn:SE_rhopi_unquenched}
    S_\pi = (Z_{\pi N} + Z_{\pi \pi N}) \log (P_0^+ V) - (1 - Z_{\pi N} - Z_{\pi \pi N}) \log (1 - Z_{\pi N} - Z_{\pi \pi N}) \\
    - \int \dd x \int \frac{\dd^2 k_{\perp}}{(2\pi)^3} f_N^{(\pi^n N)}(x,  {k}_{\perp}) \log f_N^{(\pi^n N)}(x,  {k}_{\perp}).
\end{multline}
Figure~\ref{fig:EE_unquenched} shows $S_N$, $S_\pi$, and $S_{\bar N}$ as functions of the coupling $\alpha$, alongside the quenched entropy $S_E^{\text{Que}}$ for comparison. All entropies increase with $\alpha$, except $S_{\bar N}$, which remains small due to the suppressed $N N \bar N$ component. Unlike in the quenched theory, $S_N \ne S_\pi$ in general, but they remain close -- confirming that the pion cloud dominates the entanglement structure, and the quenched approximation is quantitatively reliable.

\begin{figure}
\centering
\includegraphics[width=0.47\textwidth]{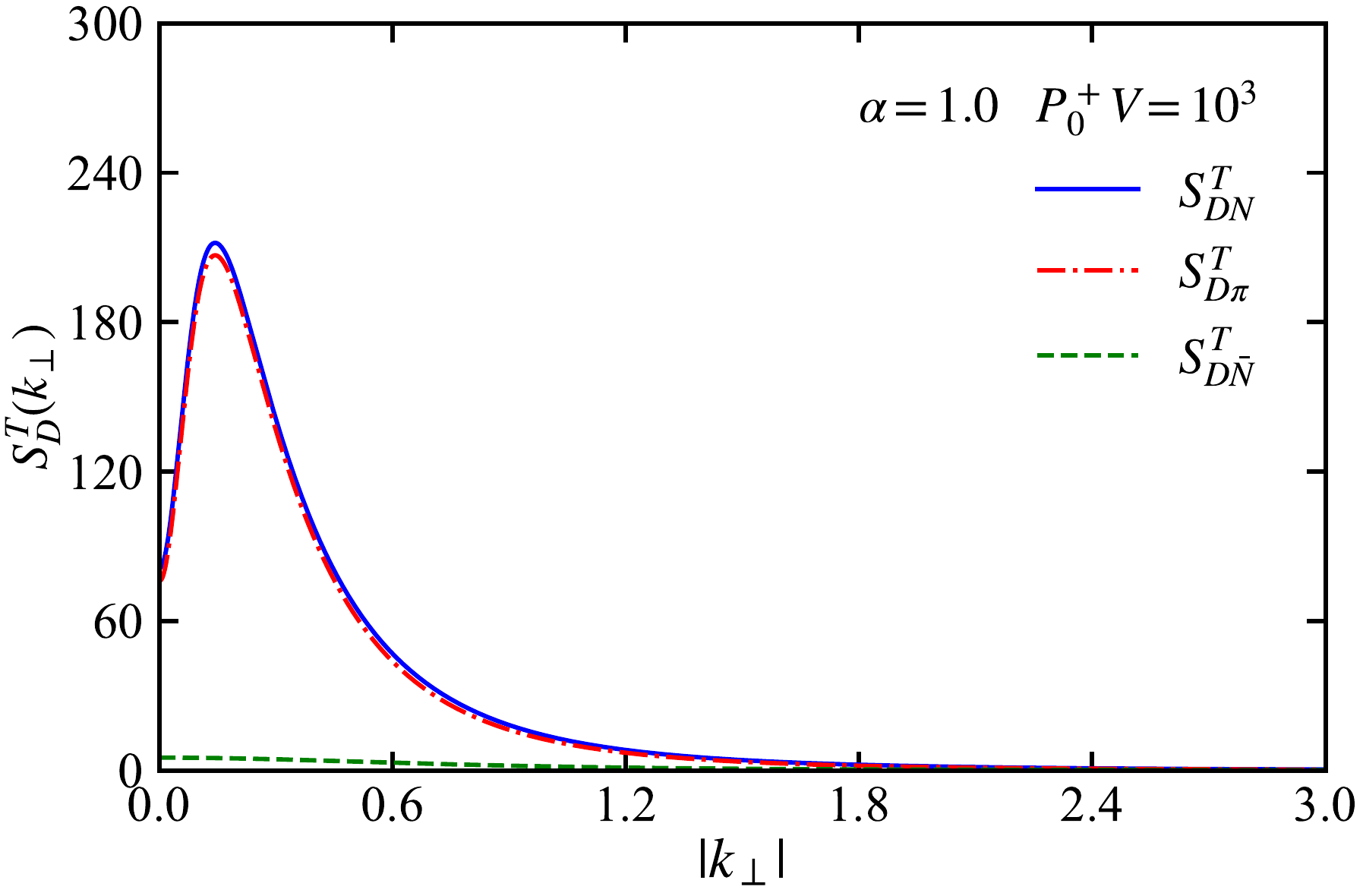}
\includegraphics[width=0.47\textwidth]{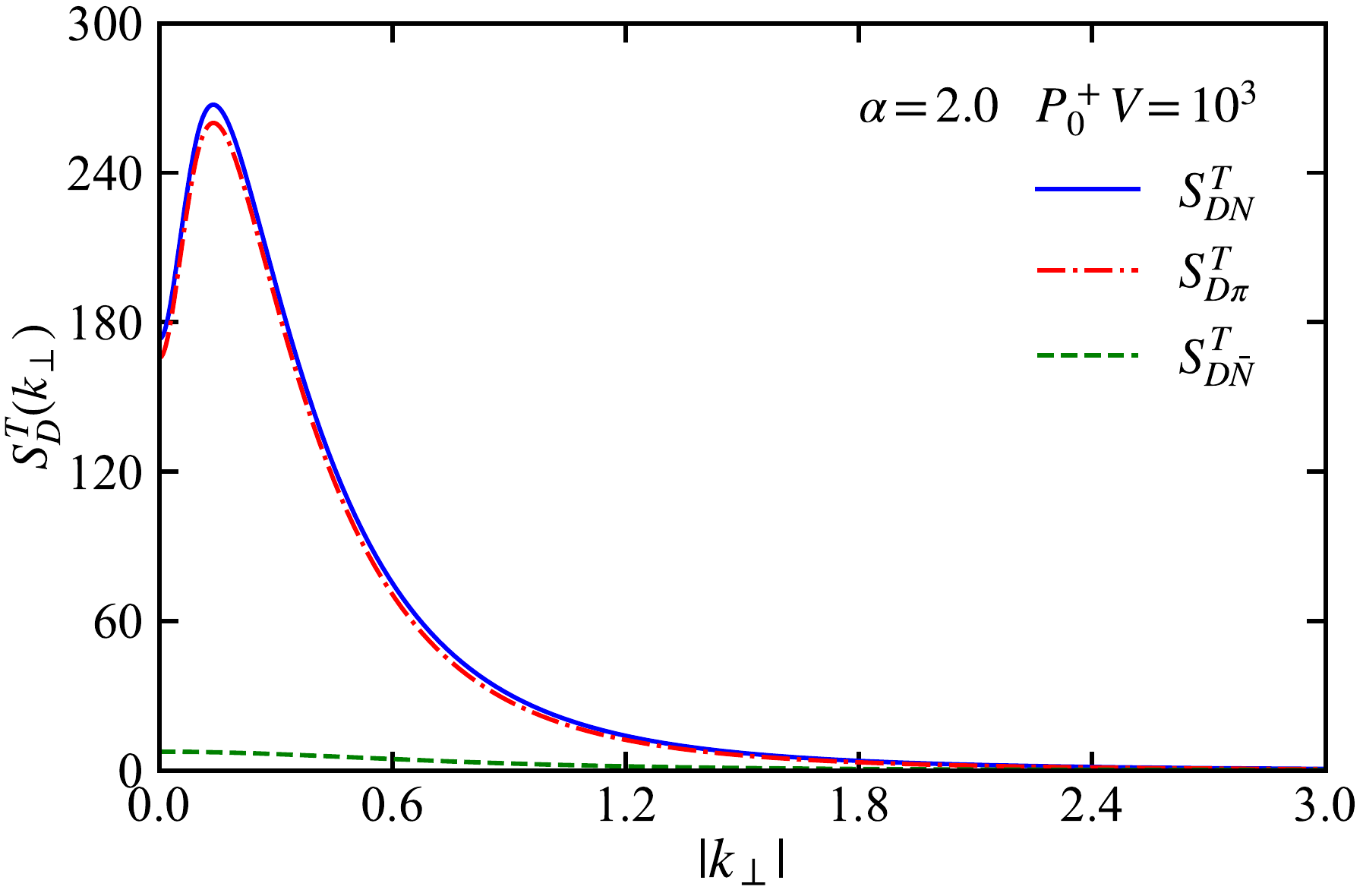}
\includegraphics[width=0.47\textwidth]{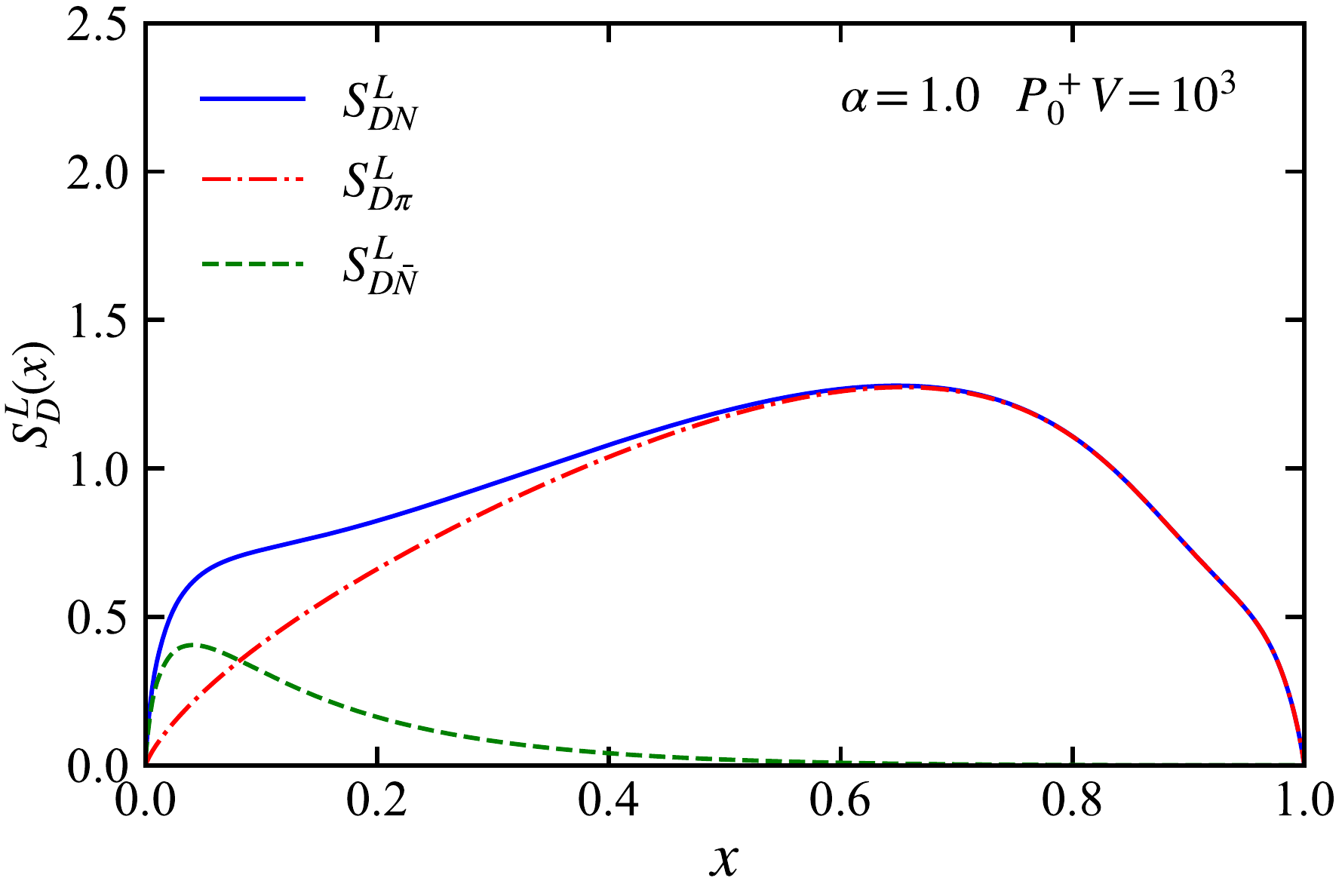}
\includegraphics[width=0.47\textwidth]{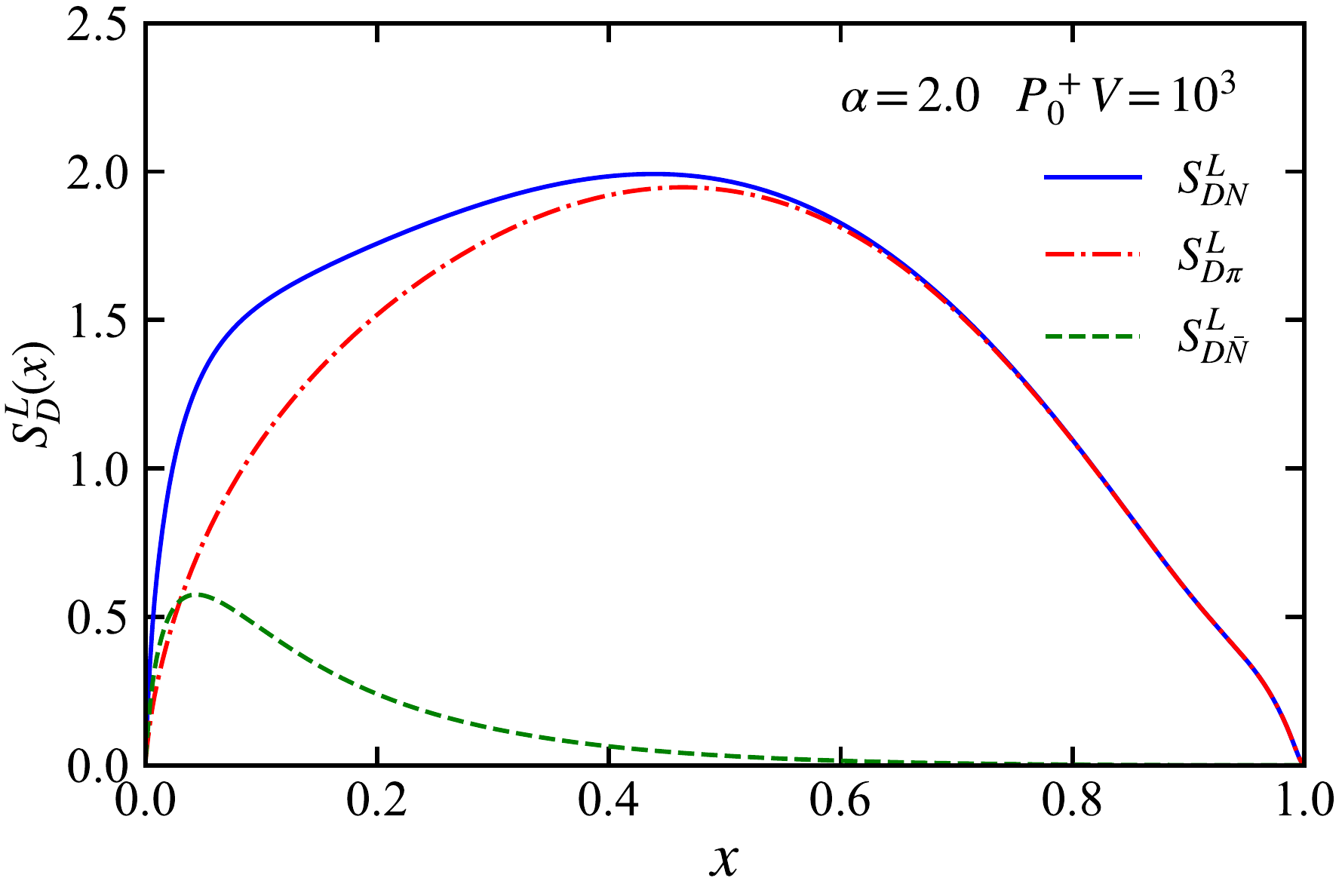}
\caption{Entropy densities in transverse (\textit{top}) and longitudinal (\textit{bottom}) momentum for $\alpha = 1.0$ (\textit{left}) and $\alpha = 2.0$ (\textit{right}). All units are in GeV.}
\label{fig:EE_density_unquenched}
\end{figure}

The momentum-space distribution of entanglement is shown in Fig.~\ref{fig:EE_density_unquenched}. As in the quenched case, the transverse entropy density peaks near ${k}_\perp \approx 0.1\,\mathrm{GeV}$, around the pion mass scale. In the longitudinal direction, $S_{DN}^L(x)$ exhibits a broad maximum at moderate $x$, with a developing secondary peak at small $x$, a signature of anti-nucleon contributions.

Entanglement entropy in the unquenched theory quantifies the entanglement between a specific parton species (e.g., the nucleon $N$) and the rest of the system, but it does not directly probe the entanglement between distinct parton species, such as $N$ and $\pi$. 
To characterize correlations between two subsystems, the mutual information is a more suitable measure. In the quenched theory, the system is bipartite, comprising only the nucleon and the pion, and the mutual information reduces to twice the entanglement entropy:
\begin{equation}
    I(N\!:\!\pi) = S_\text{vN}(\rho_N) + S_\text{vN}(\rho_\pi) - S_\text{vN}(\rho) = 2S_E^{\text{Que}},
\end{equation}
where $S_E^{\text{Que}} = S_\text{vN}(\rho_N) = S_\text{vN}(\rho_\pi)$, given that the global state is pure.
For a tripartite system with Hilbert space $\mathcal H = \mathcal H_A \otimes \mathcal H_B \otimes \mathcal H_C$, as in our unquenched model, the mutual information between subsystems $A$ and $B$ is defined as
\begin{equation}
    I(A\!:\!B) = S_\text{vN}(\rho_A) + S_\text{vN}(\rho_B) - S_\text{vN}(\rho_{AB}) = S_\text{vN}(\rho_A) + S_\text{vN}(\rho_B) - S_\text{vN}(\rho_{C}),
\end{equation}
where we used the purity of the global state to equate $S_\text{vN}(\rho_{AB}) = S_\text{vN}(\rho_C)$.

Applying this to our partonic decomposition ($A = \pi$, $B = N$, $C = \bar N$), we obtain the mutual informations between distinct parton species:
\begin{equation}\label{eq:IpiN}
\begin{split}
    I(\pi : N)&=S_\text{vN}(\rho_{\pi}) + S_\text{vN}(\rho_{N}) - S_\text{vN}(\rho_{\bar N})\\
    &=2(Z_{\pi N}+Z_{\pi\pi N}) \log (P_0^+V)-(Z+Z_{NN\bar{N}}) \log (Z+Z_{NN\bar{N}})-Z \log Z\\
    &+(1-Z_{NN\bar{N}}) \log (1-Z_{NN\bar{N}})-2\int \dd x\int\frac{\dd^2k_{\perp}}{(2\pi)^3} f_N^{(\pi^nN)}(x,k_{\perp}) \log f_N^{(\pi^nN)}(x,k_{\perp});
\end{split}
\end{equation}
\begin{equation}\label{eq:INNbar}
\begin{split}
    I(N : \bar{N})&=S_\text{vN}(\rho_{N}) + S_\text{vN}(\rho_{\bar{N}}) - S_\text{vN}(\rho_{\pi})\\
    &=2Z_{NN\bar{N}} \log (P_0^+V)-Z \log Z-(1-Z_{NN\bar{N}}) \log (1-Z_{NN\bar{N}})\\
    &+(Z+Z_{NN\bar{N}}) \log (Z+Z_{NN\bar{N}})-2\int\dd x\int\frac{\dd^2k_{\perp}}{(2\pi)^3}f_{\bar N}(x,k_{\perp}) \log f_{\bar N}(x,k_{\perp});
\end{split}
\end{equation}
\begin{equation}\label{eq:IpiNbar}
\begin{split}
    I(\pi : \bar{N})& = S_\text{vN}(\rho_{\pi}) + S_\text{vN}(\rho_{\bar{N}}) - S_\text{vN}(\rho_{N})\\
    &=-(Z+Z_{NN\bar{N}}) \log (Z+Z_{NN\bar{N}})-(1-Z_{NN\bar{N}}) \log (1-Z_{NN\bar{N}})+Z \log Z\\
    &-2\int \dd x \int\frac{\dd^2k_{\perp}}{(2\pi)^3} \Big\{ f_{N}^{(\pi^n N)}(x,k_{\perp}) \log f_{N}^{(\pi^n N)}(x,k_{\perp}) + f_{\bar N}(x,k_{\perp}) \log f_{\bar N}(x,k_{\perp}) \Big\}.
\end{split}
\end{equation}

\begin{figure}
    \centering
    \includegraphics[width=0.5\textwidth]{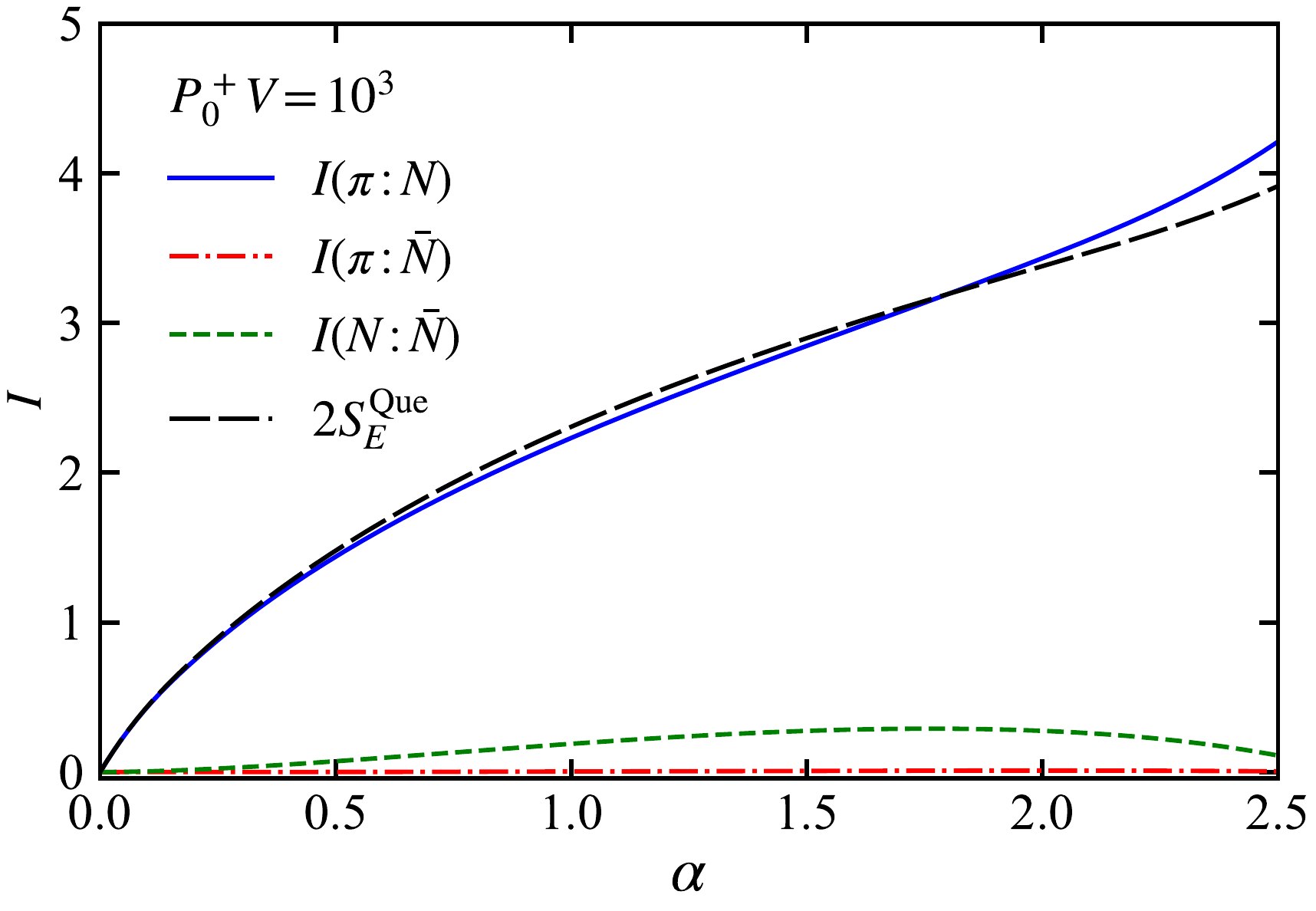}\hfill
    \caption{Mutual information between parton species as a function of the coupling $\alpha$ in the unquenched three-body theory. The quenched result $I(\pi\!:\!N) = 2S_E^{\text{Que}}$ is shown for reference.}
    \label{fig:mutual_information_unquenched}
\end{figure}

Figure~\ref{fig:mutual_information_unquenched} displays these mutual informations as functions of the coupling $\alpha$. For comparison, we also show the quenched result $I(\pi\!:\!N) = 2S_E^{\text{Que}}$. The nucleon-pion mutual information $I(\pi\!:\!N)$ remains large and close to the quenched value, confirming that the pion cloud dominates the nucleon's quantum correlations. In contrast, both $I(N\!:\!\bar N)$ and $I(\pi\!:\!\bar N)$ are small, reflecting the negligible weight of the $\bar N$ component in the Fock expansion. Notably, $I(\pi\!:\!\bar N)$ is nearly zero -- consistent with the three-body truncation, which excludes Fock sectors containing both pions and anti-nucleons simultaneously (e.g., $|{\pi N N \bar N}\rangle$ is not included beyond $|{N N \bar N}\rangle$).

\begin{figure}
    \centering
    \includegraphics[width=0.47\textwidth]{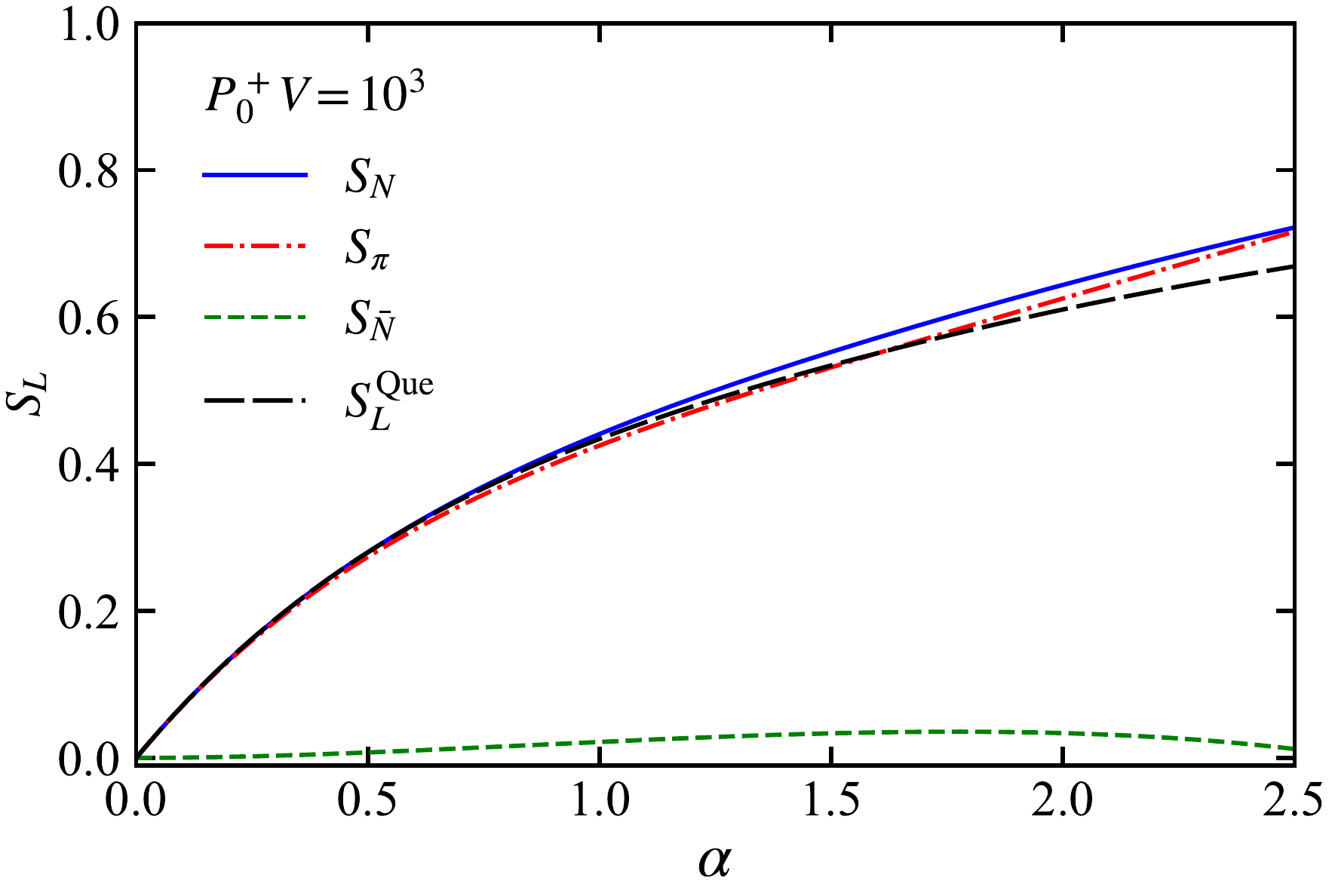}\hfill
    \includegraphics[width=0.47\textwidth]{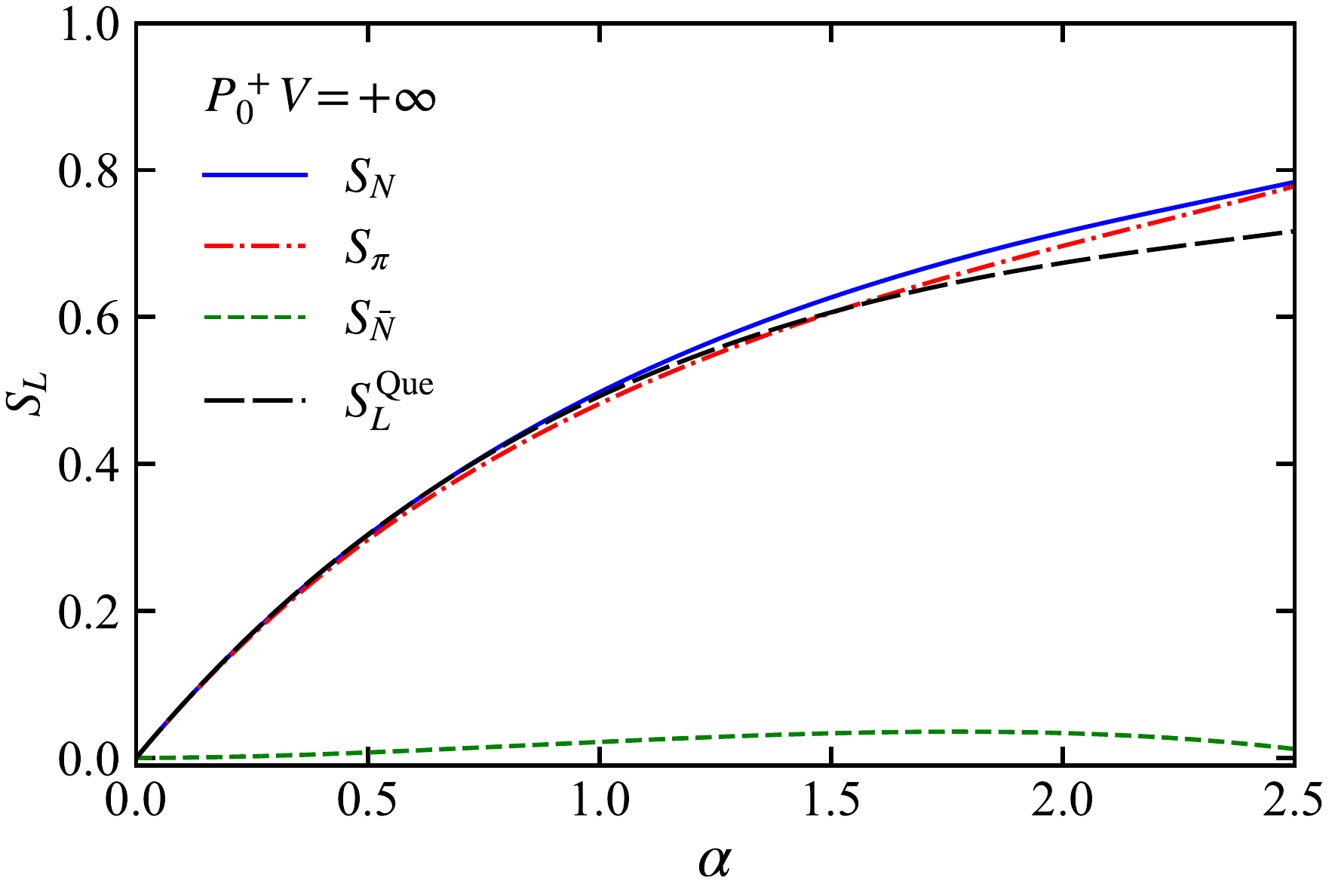}
    \caption{Linear entropy as a function of $\alpha$ in the unquenched three-body theory: (\textit{left}) finite volume $P_0^+ V = 10^3\,\mathrm{GeV}^{-2}$; (\textit{right}) continuum limit $V \to \infty$.}
    \label{fig:linear_entropy_unquenched}
\end{figure}

Finally, as an alternative entanglement witness, we compute the linear entropy $S_L = 1 - \operatorname{Tr}(\rho^2)$ for each subsystem. Figure~\ref{fig:linear_entropy_unquenched} shows $S_L$ for both finite volume ($P_0^+ V = 10^3\,\mathrm{GeV}^{-2}$) and the continuum limit ($V \to \infty$), illustrating its convergence behavior with coupling strength.

\section{Discussions}\label{sec:discussions}

\begin{figure}
    \centering
    \includegraphics[width=0.47\textwidth]{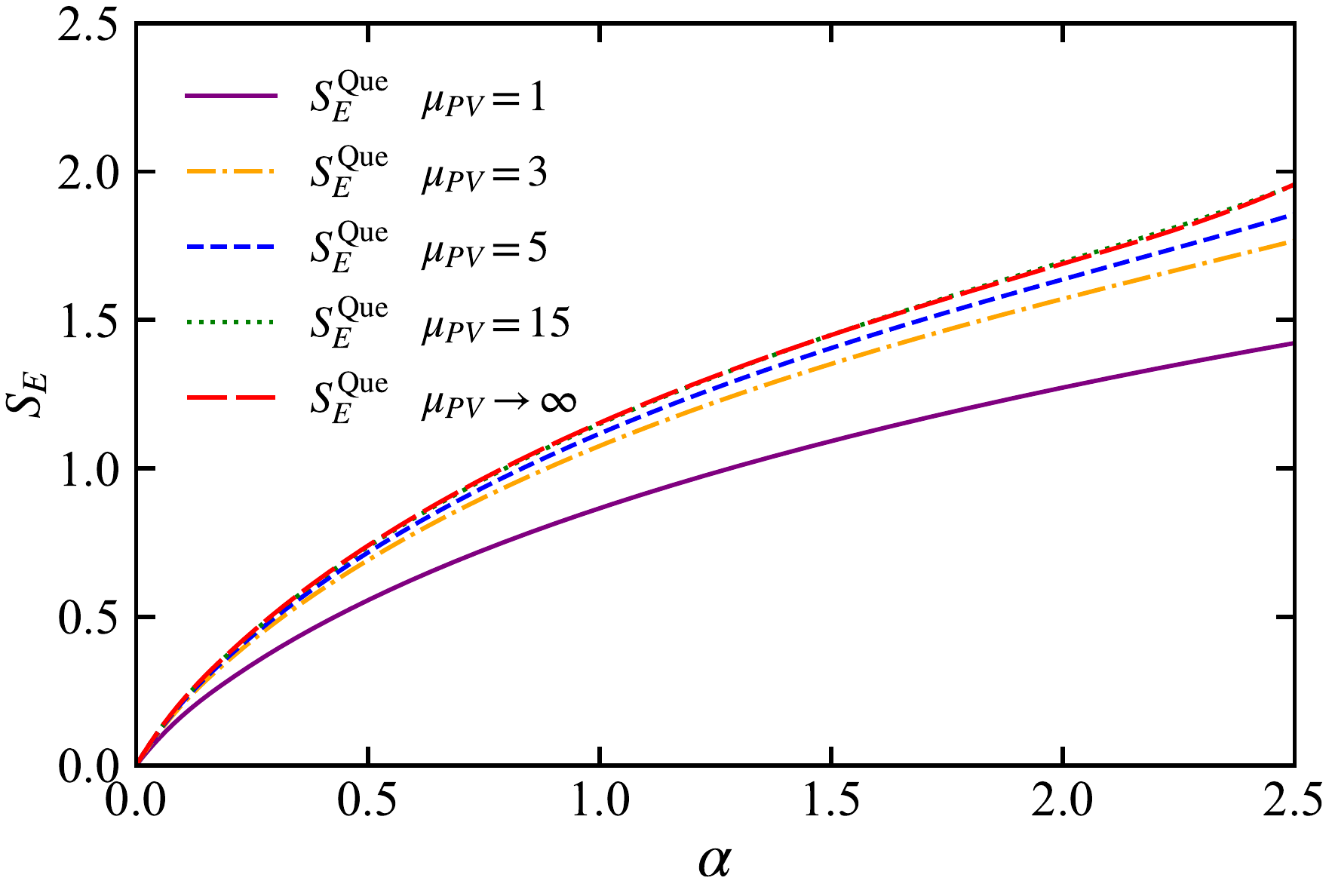}\hfill
    \includegraphics[width=0.47\textwidth]{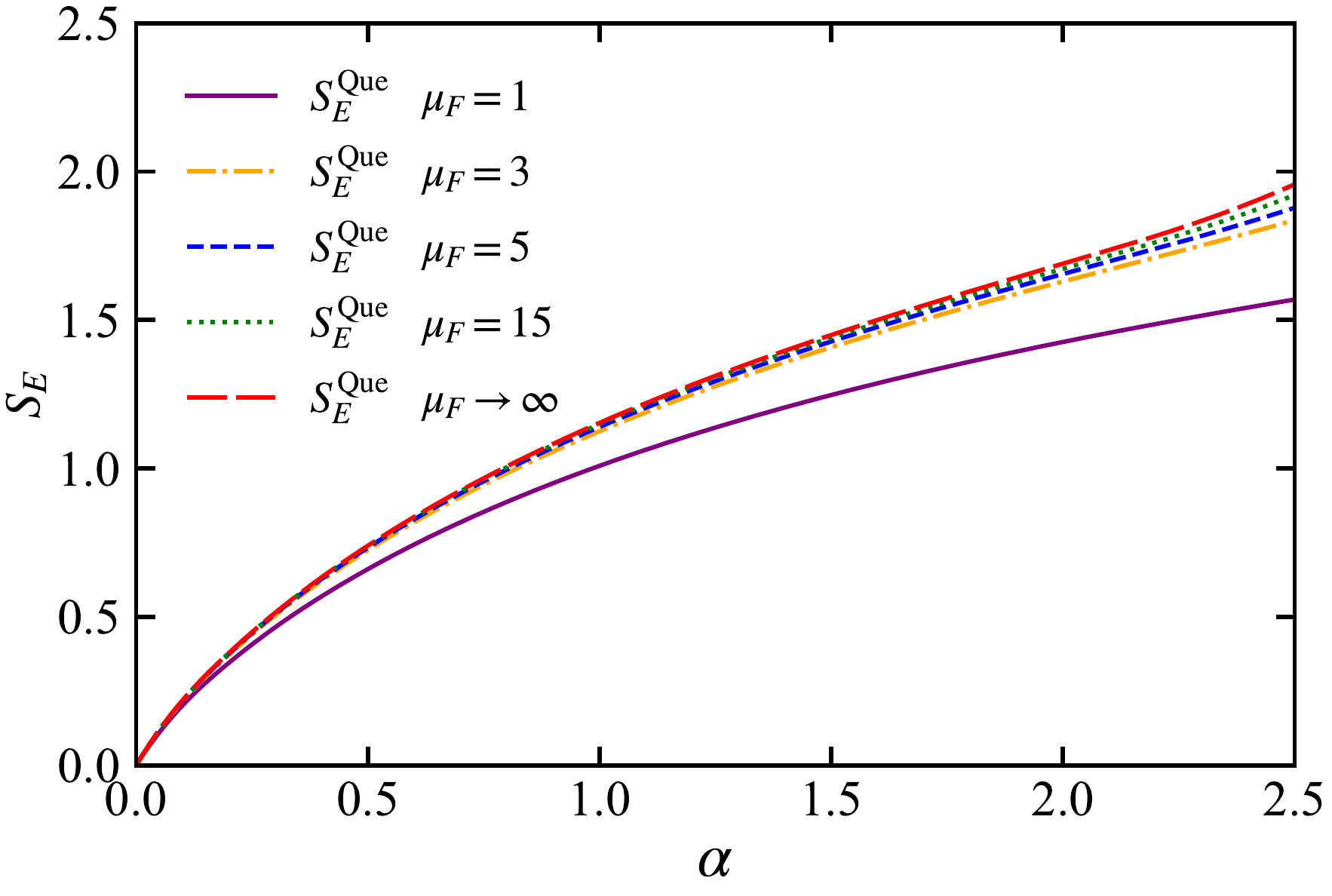}
    \caption{Scale dependence of the entanglement entropy in the quenched theory. (\textit{Left}): dependence on the UV regulator $\mu_\text{PV}$. (\textit{Right}): dependence on the factorization scale $\mu_F$, which acts as a transverse momentum cutoff in defining the reduced density matrix.}
    \label{fig:scale_dependence}
\end{figure}

Like other observables in QFT, the entanglement entropy depends on UV cutoff. In the scalar Yukawa model, we employ PV regularization, where the PV mass $\mu_\text{PV}$ acts as a UV regulator. Owing to the model’s super-renormalizability, its UV structure is particularly simple. As shown in the left panel of Fig.~\ref{fig:scale_dependence}, the quenched entanglement entropy $S_E^{\text{Que}}$ converges in the limit $\mu_\text{PV} \to \infty$. Consequently, we have omitted the UV regulator in all preceding calculations.

In experiments, the transverse momentum of a parton is only observable within the finite acceptance of the detector. Therefore, the density matrix, and hence the entanglement entropy, is only accessible up to a finite factorization scale $\mu_F$. Modes with transverse momentum $k_\perp \gtrsim \mu_F$ are unresolved and must be traced out. This motivates the definition of a scale-dependent reduced density matrix:
\begin{equation}
    \tilde \rho_N = \operatorname{Tr}_{k_\perp > \mu_F} \rho_N.
\end{equation}
The corresponding entanglement entropy in the quenched theory becomes
\begin{multline}
    S_{\text{vN}}(\tilde \rho_N) = (1 - Z_F) \log (P_0^+ V) - Z_F \log Z_F \\
    - \int \dd x \int_{k_{\perp} \le \mu_F} \frac{\dd^2 k_{\perp}}{(2\pi)^3}
    \sum_{i \geq 2} f_N^{(i)}(x, {k}_{\perp}) \log \Bigg[ \sum_{i \geq 2} f_N^{(i)}(x, {k}_{\perp}) \Bigg],
\end{multline}
where the effective field-strength renormalization constant at scale $\mu_F$ is
\begin{equation}
    Z_F = 1 - \int \dd x \int_{k_\perp \le \mu_F} \frac{\dd^2 k_{\perp}}{(2\pi)^3} \sum_{i \geq 2} f_N^{(i)}(x, \vec{k}_{\perp}).
\end{equation}
The right panel of Fig.~\ref{fig:scale_dependence} displays this scale dependence. Analogous to the UV behavior, $S_E^{\text{Que}}$ remains finite as $\mu_F \to \infty$, again reflecting the simplicity of the scalar theory’s UV structure.
Of course, in QCD, the scale dependence of entanglement entropy is expected to be highly nontrivial due to asymptotic freedom, operator mixing, and rapidity divergences, and would require careful treatment beyond the present model \cite{Neill:2018uqw}.

Our results contain an IR divergence $\log P^+_0V$. This divergence naturally splits into longitudinal and transverse parts:
\[
\log (P^+_0V) = \log (2\pi K) + \log (L^2_\perp),
\]
where $K = L P^+_0/(2\pi)$ is the light-cone harmonic resolution. The longitudinal contribution satisfies the area law characteristic of 1+1D quantum field theories, as demonstrated in Ref.~\cite{Liu:2022ohy}. In that work, the quantity $\chi = \log(P^+_0 L) = \log(2\pi K)$ is identified with rapidity, providing a direct link between entanglement entropy and longitudinal kinematics. 
This connection is reinforced by broader studies showing that, in the high-energy limit, entanglement entropy grows linearly with rapidity -- a universal feature observed in holographic models~\cite{Gursoy:2023hge}, two-dimensional QCD~\cite{Liu:2022ohy}, and early theoretical analyses of parton dynamics~\cite{Kharzeev:2017qzs}.

An intriguing question concerns the maximum possible value of the entropy functional \cite{Asadi:2022vbl, Asadi:2023bat}:
\begin{equation}
    H(f) = \log (P^+_0 V) - \int \dd x \int \frac{\dd^2 k_\perp}{(2\pi)^3} f(x,  {k}_\perp) \log f(x,  {k}_\perp)?
\end{equation}
If $f(x, {k}_\perp)$ is constrained only by normalization,
\begin{equation}
    \int \dd x \int \frac{\dd^2 k_\perp}{(2\pi)^3} f(x,  {k}_\perp) = 1,
\end{equation}
the entropy is maximized by a uniform distribution over the available phase space. Introducing a transverse momentum cutoff $\Lambda$, this yields
\begin{equation}
    f(x,  {k}_\perp) = \frac{(2\pi)^3}{\Lambda^2},
\end{equation}
and the corresponding maximal entropy is
\begin{equation}
    S_\text{max} = \log P_0^+ V + \log \frac{\Lambda^2}{(2\pi)^3}.
\end{equation}
Comparing to our results, the entanglement entropy is far from saturating the maximal entropy. 
If, in addition, $f(x, {k}_\perp)$ is subject to an energy constraint, such as those arising in light-front holography \cite{Brodsky:2014yha},
\begin{equation}
    \int \dd x \int \frac{\dd^2 k_\perp}{(2\pi)^3} f(x,  {k}_\perp)\mathcal{M}^2(x, k_\perp) = M^2,
\end{equation}
where $\mathcal M^2 = \frac{k_\perp^2 + m^2}{x} +  \frac{k_\perp^2 + \mu^2 }{1-x} + U_\text{eff}$ is the effective light-cone Hamiltonian, 
the entropy is maximized by a Boltzmann distribution,
\begin{equation}
    f(x,  {k}_\perp) = Z^{-1} \exp\!\left( -\beta \mathcal M^2(x, k_\perp) \right),
\end{equation}
where the partition function $Z$ and inverse temperature $\beta$ are fixed by the normalization and energy constraints.

\section{Summary and outlook}
\label{sec:summary}

In this work, we performed a first-principles, non-perturbative computation of quantum entanglement between partonic constituents in a strongly coupled 3+1D scalar Yukawa theory using light-front Hamiltonian methods. By explicitly constructing reduced density matrices from exact light-front wave functions (LFWFs) and evaluating von Neumann entropies, mutual information, and linear entropy, we establish a formal link between quantum information theory and parton structure in quantum field theory. This approach provides a controlled setting to explore how entanglement encodes the non-perturbative dynamics of hadron formation, offering insights relevant to QCD and future collider phenomenology.

Our key finding is that the entanglement entropy is closely related to the Shannon entropy of the transverse momentum dependent distribution. In particular, in the quenched approximation (where pair creation in sea is omitted), the entanglement entropy exactly coincides with the Shannon entropy of the TMD. This result arises because tracing over the unobserved d.o.f.~yields a diagonal reduced density matrix thanks to the kinematical nature of the light-front boosts. In contrast, in the unquenched theory, the entanglement entropy cannot be expressed as the Shannon entropy of the TMD, original or normalized, revealing that quantum correlations encode genuinely non-classical information beyond what is accessible through standard parton distributions. 
We also examined alternative entanglement witnesses, the mutual information and the linear entropy. 
Mutual information confirms that nucleon-pion correlations dominate, while anti-nucleon contributions remain negligible in the three-body truncation, highlighting the physical relevance of the quenched approximation itself.

This correspondence between entanglement entropy and TMD Shannon entropy has important implications for effective descriptions of hadron structure. In low-resolution models, such as constituent quark approaches, where only valence d.o.f.~are retained, the quenched picture is well justified, and the Shannon entropy of parton distributions may serve as a faithful proxy for quantum entanglement. Similarly, at large momentum fraction-$x$, where sea and gluon contributions are kinematically suppressed, the valence sector dominates, and experimental access to TMDs in this regime (e.g., at the Electron-Ion Collider) could indirectly probe the underlying quantum entanglement structure of the nucleon.

Extending this framework to full QCD introduces several conceptual and technical challenges. First, we need to incorporate the spin degrees of freedom and expand our inventory to include polarized TMDs \cite{Angeles-Martinez:2015sea}. Second, gauge invariance necessitates the inclusion of Wilson lines in TMD definitions, which entangle color and kinematic d.o.f. Third, UV and rapidity divergences, along with collinear singularities, prevent a naïve parton definition: a quark or gluon observed in deep inelastic scattering is always dressed by collinear radiation. A consistent entanglement measure must therefore incorporate these collinear modes into the definition of the probed subsystem, effectively treating jet-like objects, rather than bare partons, as the fundamental entangled d.o.f.  Addressing these issues will require combining light-front quantization with soft-collinear effective theory \cite{Bauer:2001yt} or renormalization group techniques \cite{Hentschinski:2024gaa}.

The close connection between TMDs and entanglement opens new avenues for hadron structure phenomenology \cite{Cloet:2019wre}. Entanglement entropy, mutual information, and linear entropy could serve as novel observables in semi-inclusive deep inelastic scattering (SIDIS), Drell-Yan, or $e^+e^-$ annihilation processes \cite{Qi:2025onf, Moriggi:2025qfs, Brandenburg:2025one}. In particular, large-$x$ measurements, where the quenched approximation holds, may allow extraction of entanglement signatures directly from TMD data \cite{Wang:2022noa}. Furthermore, the rapidity dependence of entanglement entropy could provide a quantum-information-theoretic interpretation of parton evolution, complementing traditional DGLAP or BFKL frameworks \cite{Qian:2021jxp, Kreshchuk:2020aiq}.

The structure of the light-front ground state encoded in its entanglement properties suggests a natural representation via tensor networks (e.g., matrix product states or projected entangled-pair states) \cite{Belyansky:2023rgh}. Entanglement entropy can serve as a key diagnostic for the bond dimension required by such representations. For example,  a logarithmic scaling would indicate a critical or conformal structure, while area-law scaling supports efficient tensor network approximations. Unlike the equal-time quantization, light-front quantized QFTs exhibit separate longitudinal and transverse scalings, indicating that light-front QFTs may be more efficiently simulated using quantum many-body methods \cite{Alterman:2025prb}. 
This perspective may facilitate quantum simulations of hadron wave functions on near-term quantum hardware. 

Finally, richer quantum information measures, such as the entanglement spectrum and logarithmic negativity, could reveal nontrivial features of strongly coupled bound states. For example, degeneracies or universal scaling in the entanglement spectrum may signal emergent symmetries or topological order, while negativity could detect entanglement in mixed-state scenarios (e.g., in high-multiplicity final states). These tools offer a pathway to uncover universal quantum signatures of confinement, chiral symmetry breaking, and hadron mass generation beyond what traditional correlation functions can provide.
Together, these directions position entanglement not merely as a theoretical curiosity, but as a fundamental diagnostic of quantum structure in QFT, with direct implications for both theoretical advances and experimental programs in high-energy physics.

\section*{Acknowledgements}

 This work was supported in part by the Chinese Academy of Sciences under Grant No.~YSBR-101, by the National Natural Science Foundation of China (NSFC) under Grant Nos.~12375081, 125B2111. 
WQ is supported by the European Research Council under project ERC-2018-ADG-835105 YoctoLHC; by Maria de Maeztu excellence unit grant CEX2023-001318-M and project PID2020-119632GB-I00 funded by MICIU/AEI/10.13039/501100011033; by ERDF/EU; by the Marie Sklodowska-Curie Actions Fellowships under Grant No.~101109293; and by Xunta de Galicia (CIGUS Network of Research Centres). 
Y.Z. is supported by the European Union ``Next Generation EU'' program through the Italian PRIN 2022 grant n. 20225ZHA7W.

\appendix

\section{Pauli-Villars regularization}\label{sec:PVreg}

In Section~\ref{sec:entanglement_in_unquenched_theory}, we noted that Pauli-Villars (PV) pion must be introduced to regularize ultraviolet divergences in the scalar Yukawa theory.
To perform numerical calculations of the entanglement entropy in the quenched scalar Yukawa theory, we must restore the PV-regulated LFWFs in all relevant expressions. Upon reintroducing the PV fields, the squared modulus of the $n$-body LFWF is replaced by an alternating sum over physical and PV pion configurations:
\begin{equation}
    |\psi_n(\{x_i,\vec{k}_{i\perp}\})|^2 \;\longrightarrow\;
    \sum_{l_1,l_2,\dots,l_{n-1}=0}^{1} (-1)^{l_1 + l_2 + \cdots + l_{n-1}}\,
    \big|\psi_n^{l_1 l_2 \dots l_{n-1}}(\{x_i,\vec{k}_{i\perp}\})\big|^2,
\end{equation}
where each index $l_a \in \{0,1\}$ labels whether the $a$-th pion is a physical pion ($l_a = 0$) or a PV  pion ($l_a = 1$).

Consequently, the nucleon TMD in Eq.~\eqref{eqn:TMD} is modified to
\begin{multline}\label{eq:PV_TMD}
f_N(x, {k}_\perp) = \sum_{n} \sum_{l_1, \cdots, l_{n-1}} (-1)^{l_1+\cdots +l_{n-1}} \int [\dd x_i \dd^2 k_{i\perp}]_n \, (2\pi)^3 \delta(x - x_n) \delta^2( {k}_\perp -  {k}_{n\perp}) \\ \times \big|\psi_n^{l_1\cdots l_{n-1}}(\{x_i, \vec{k}_{i\perp}\})\big|^2.
\end{multline}
This expression can be recast in a more compact and physically transparent form by recognizing that the alternating sum implements the PV subtraction at the level of probability densities. The resulting TMD remains positive-definite in the continuum limit after renormalization, and all UV divergences cancel order by order in the Fock expansion.


\begin{thebibliography}{99}

\bibitem{Gross:2022hyw}
F.~Gross, E.~Klempt, S.~J.~Brodsky, A.~J.~Buras, V.~D.~Burkert, G.~Heinrich, K.~Jakobs, C.~A.~Meyer, K.~Orginos and M.~Strickland, \textit{et al.}
``50 Years of Quantum Chromodynamics,''
Eur. Phys. J. C \textbf{83}, 1125 (2023)
doi:10.1140/epjc/s10052-023-11949-2
[arXiv:2212.11107 [hep-ph]].

\bibitem{ParticleDataGroup:2024cfk}
S.~Navas \textit{et al.} [Particle Data Group],
``Review of particle physics,''
Phys. Rev. D \textbf{110}, no.3, 030001 (2024)
doi:10.1103/PhysRevD.110.030001

\bibitem{Kharzeev:2017qzs}
D.~E.~Kharzeev and E.~M.~Levin,
``Deep inelastic scattering as a probe of entanglement,''
Phys. Rev. D \textbf{95}, no.11, 114008 (2017)
doi:10.1103/PhysRevD.95.114008
[arXiv:1702.03489 [hep-ph]].

\bibitem{Kharzeev:2021nzh}
D.~E.~Kharzeev,
``Quantum information approach to high energy interactions,''
Phil. Trans. A. Math. Phys. Eng. Sci. \textbf{380}, no.2216, 20210063 (2021)
doi:10.1098/rsta.2021.0063
[arXiv:2108.08792 [hep-ph]].

\bibitem{Nishioka:2018khk}
T.~Nishioka,
``Entanglement entropy: holography and renormalization group,''
Rev. Mod. Phys. \textbf{90}, no.3, 035007 (2018)
doi:10.1103/RevModPhys.90.035007
[arXiv:1801.10352 [hep-th]].

\bibitem{Witten:2018zxz}
E.~Witten,
``APS Medal for Exceptional Achievement in Research: Invited article on entanglement properties of quantum field theory,''
Rev. Mod. Phys. \textbf{90}, no.4, 045003 (2018)
doi:10.1103/RevModPhys.90.045003
[arXiv:1803.04993 [hep-th]].

\bibitem{Liu:2018gae}
Y.~Liu and I.~Zahed,
``Entanglement in Regge scattering using the AdS/CFT correspondence,''
Phys. Rev. D \textbf{100}, no.4, 046005 (2019)
doi:10.1103/PhysRevD.100.046005
[arXiv:1803.09157 [hep-ph]].

\bibitem{Beane:2019loz}
S.~R.~Beane and P.~Ehlers,
``Chiral symmetry breaking, entanglement, and the nucleon spin decomposition,''
Mod. Phys. Lett. A \textbf{35}, no.08, 2050048 (2019)
doi:10.1142/S0217732320500480
[arXiv:1905.03295 [hep-ph]].

\bibitem{Feal:2020myr}
X.~Feal, C.~Pajares and R.~Vazquez,
``Thermal and hard scales in transverse momentum distributions, fluctuations, and entanglement,''
Phys. Rev. C \textbf{104}, no.4, 044904 (2021)
doi:10.1103/PhysRevC.104.044904
[arXiv:2012.02894 [hep-ph]].

\bibitem{Liu:2022ohy}
Y.~Liu, M.~A.~Nowak and I.~Zahed,
``Entanglement entropy and flow in two-dimensional QCD: Parton and string duality,''
Phys. Rev. D \textbf{105}, no.11, 114027 (2022)
doi:10.1103/PhysRevD.105.114027
[arXiv:2202.02612 [hep-ph]].

\bibitem{Liu:2022hto}
Y.~Liu, M.~A.~Nowak and I.~Zahed,
``Rapidity evolution of the entanglement entropy in quarkonium: Parton and string duality,''
Phys. Rev. D \textbf{105}, no.11, 114028 (2022)
doi:10.1103/PhysRevD.105.114028
[arXiv:2203.00739 [hep-ph]].

\bibitem{Liu:2022qqf}
Y.~Liu, M.~A.~Nowak and I.~Zahed,
``Spatial entanglement in two-dimensional QCD: Renyi and Ryu-Takayanagi entropies,''
Phys. Rev. D \textbf{107}, no.5, 054010 (2023)
doi:10.1103/PhysRevD.107.054010
[arXiv:2205.06724 [hep-ph]].

\bibitem{Ehlers:2022oke}
P.~J.~Ehlers,
``Entanglement between Quarks in Hadrons,''
Ph.D. thesis, Washington U., Seattle (2022)

\bibitem{Liu:2023zno}
Y.~Liu, M.~A.~Nowak and I.~Zahed,
``Nambu-Goto string in QCD: Dipole interactions, scattering, and entanglement,''
Phys. Rev. D \textbf{108}, no.9, 094025 (2023)
doi:10.1103/PhysRevD.108.094025
[arXiv:2301.06154 [hep-ph]].

\bibitem{Grieninger:2023ufa}
S.~Grieninger, K.~Ikeda, D.~E.~Kharzeev and I.~Zahed,
``Entanglement in massive Schwinger model at finite temperature and density,''
Phys. Rev. D \textbf{109}, no.1, 016023 (2024)
doi:10.1103/PhysRevD.109.016023
[arXiv:2312.03172 [hep-th]].

\bibitem{Florio:2023mzk}
A.~Florio,
``Two-fermion negativity and confinement in the Schwinger model,''
Phys. Rev. D \textbf{109}, no.7, L071501 (2024)
doi:10.1103/PhysRevD.109.L071501
[arXiv:2312.05298 [hep-th]].

\bibitem{Ozzello:2025tfu}
Z.~Ozzello and Y.~Meurice,
``Multipartite entanglement from ditstrings for 1+1D systems,''
[arXiv:2507.14422 [quant-ph]].

\bibitem{Florio:2025xup}
A.~Florio and S.~Murciano,
``Entanglement asymmetry in gauge theories: chiral anomaly in the finite temperature massless Schwinger model,''
[arXiv:2511.01966 [hep-th]].

\bibitem{CarrascoMillan:2018ufj}
P.~Carrasco Mill{\'a}n, M.~{\'A}.~Garc{\'\i}a-Ferrero, F.~J.~Llanes-Estrada, A.~Porras Riojano and E.~M.~S{\'a}nchez Garc{\'\i}a,
Nucl. Phys. B \textbf{930}, 583-596 (2018)
doi:10.1016/j.nuclphysb.2018.04.003
[arXiv:1802.05487 [hep-ph]].

\bibitem{Benito-Calvino:2022kqa}
G.~Benito-Calvi{\~n}o, J.~Garc{\'\i}a-Olivares and F.~J.~Llanes-Estrada,
Nucl. Phys. A \textbf{1036}, 122670 (2023)
doi:10.1016/j.nuclphysa.2023.122670
[arXiv:2209.13225 [hep-ph]].

\bibitem{Galvez-Viruet:2025rmy}
J.~J.~G{\'a}lvez-Viruet, F.~J.~Llanes-Estrada, N.~M.~de Arenaza, M.~G{\'o}mez-Rocha and T.~J.~Hobbs,
[arXiv:2510.18869 [hep-ph]].

\bibitem{Dumitru:2022tud}
A.~Dumitru and E.~Kolbusz,
``Quark and gluon entanglement in the proton on the light cone at intermediate $x$,''
Phys. Rev. D \textbf{105}, 074030 (2022)
doi:10.1103/PhysRevD.105.074030
[arXiv:2202.01803 [hep-ph]].

\bibitem{Dumitru:2023fih}
A.~Dumitru and E.~Kolbusz,
``Quark pair angular correlations in the proton: Entropy versus entanglement negativity,''
Phys. Rev. D \textbf{108}, no.3, 034011 (2023)
doi:10.1103/PhysRevD.108.034011
[arXiv:2303.07408 [hep-ph]].

\bibitem{Dumitru:2023qee}
A.~Dumitru, A.~Kovner and V.~V.~Skokov,
``Entanglement entropy of the proton in coordinate space,''
Phys. Rev. D \textbf{108}, no.1, 014014 (2023)
doi:10.1103/PhysRevD.108.014014
[arXiv:2304.08564 [hep-ph]].

\bibitem{Dosch:2023bxj}
H.~G.~Dosch, G.~F.~de Teramond and S.~J.~Brodsky,
``Entropy from entangled parton states and high-energy scattering behavior,''
Phys. Lett. B \textbf{850}, 138521 (2024)
doi:10.1016/j.physletb.2024.138521
[arXiv:2304.14207 [hep-ph]].

\bibitem{Qian:2024fqf}
C.~Qian, S.~Xu, Y.~G.~Yang and X.~Zhao,
``Quark and gluon entanglement in the proton based on a light-front Hamiltonian,''
Phys. Rev. D \textbf{112}, no.1, 014004 (2025)
doi:10.1103/wlm7-x1wn
[arXiv:2412.11860 [hep-ph]].

\bibitem{Dumitru:2025bib}
A.~Dumitru and E.~Kolbusz,
``Quantum entanglement correlations in double quark PDFs,''
Phys. Rev. D \textbf{111}, no.11, 114033 (2025)
doi:10.1103/ngdx-pc85
[arXiv:2501.12312 [hep-ph]].

\bibitem{Kolbusz:2025vhh}
E.~Kolbusz,
``Quantum Entanglement Correlations in the Proton on the Light-Front,''
Ph.D. thesis,  CUNY, Graduate School - U. Ctr. (2025)

\bibitem{Hagiwara:2017uaz}
Y.~Hagiwara, Y.~Hatta, B.~W.~Xiao and F.~Yuan,
``Classical and quantum entropy of parton distributions,''
Phys. Rev. D \textbf{97}, no.9, 094029 (2018)
doi:10.1103/PhysRevD.97.094029
[arXiv:1801.00087 [hep-ph]].

\bibitem{Kovner:2018rbf}
A.~Kovner, M.~Lublinsky and M.~Serino,
``Entanglement entropy, entropy production and time evolution in high energy QCD,''
Phys. Lett. B \textbf{792}, 4-15 (2019)
doi:10.1016/j.physletb.2018.10.043
[arXiv:1806.01089 [hep-ph]].

\bibitem{Peschanski:2019yah}
R.~Peschanski and S.~Seki,
``Evaluation of Entanglement Entropy in High Energy Elastic Scattering,''
Phys. Rev. D \textbf{100}, no.7, 076012 (2019)
doi:10.1103/PhysRevD.100.076012
[arXiv:1906.09696 [hep-th]].

\bibitem{Armesto:2019mna}
N.~Armesto, F.~Dominguez, A.~Kovner, M.~Lublinsky and V.~Skokov,
``The Color Glass Condensate density matrix: Lindblad evolution, entanglement entropy and Wigner functional,''
JHEP \textbf{05}, 025 (2019)
doi:10.1007/JHEP05(2019)025
[arXiv:1901.08080 [hep-ph]].

\bibitem{Tu:2019ouv}
Z.~Tu, D.~E.~Kharzeev and T.~Ullrich,
``Einstein-Podolsky-Rosen Paradox and Quantum Entanglement at Subnucleonic Scales,''
Phys. Rev. Lett. \textbf{124}, no.6, 062001 (2020)
doi:10.1103/PhysRevLett.124.062001
[arXiv:1904.11974 [hep-ph]].

\bibitem{Duan:2020jkz}
H.~Duan, C.~Akkaya, A.~Kovner and V.~V.~Skokov,
``Entanglement, partial set of measurements, and diagonality of the density matrix in the parton model,''
Phys. Rev. D \textbf{101}, no.3, 036017 (2020)
doi:10.1103/PhysRevD.101.036017
[arXiv:2001.01726 [hep-ph]].

\bibitem{Gotsman:2020bjc}
E.~Gotsman and E.~Levin,
``High energy QCD: multiplicity distribution and entanglement entropy,''
Phys. Rev. D \textbf{102}, no.7, 074008 (2020)
doi:10.1103/PhysRevD.102.074008
[arXiv:2006.11793 [hep-ph]].

\bibitem{Ramos:2020kaj}
G.~S.~Ramos and M.~V.~T.~Machado,
``Investigating entanglement entropy at small-x in DIS off protons and nuclei,''
Phys. Rev. D \textbf{101}, no.7, 074040 (2020)
doi:10.1103/PhysRevD.101.074040
[arXiv:2003.05008 [hep-ph]].

\bibitem{Kharzeev:2021yyf}
D.~E.~Kharzeev and E.~Levin,
``Deep inelastic scattering as a probe of entanglement: Confronting experimental data,''
Phys. Rev. D \textbf{104}, no.3, L031503 (2021)
doi:10.1103/PhysRevD.104.L031503
[arXiv:2102.09773 [hep-ph]].

\bibitem{Zhang:2021hra}
K.~Zhang, K.~Hao, D.~Kharzeev and V.~Korepin,
``Entanglement entropy production in deep inelastic scattering,''
Phys. Rev. D \textbf{105}, no.1, 014002 (2022)
doi:10.1103/PhysRevD.105.014002
[arXiv:2110.04881 [quant-ph]].

\bibitem{Duan:2023zke}
H.~Duan,
``Quantum Information Perspective on High Energy Hadron Wave Function: Entanglement and Correlations.,''
Ph.D. thesis, North Carolina State U. (2023)

\bibitem{Gursoy:2023hge}
U.~G{\"u}rsoy, D.~E.~Kharzeev and J.~F.~Pedraza,
``Universal rapidity scaling of entanglement entropy inside hadrons from conformal invariance,''
Phys. Rev. D \textbf{110}, no.7, 074008 (2024)
doi:10.1103/PhysRevD.110.074008
[arXiv:2306.16145 [hep-th]].

\bibitem{Datta:2024hpn}
J.~Datta, A.~Deshpande, D.~E.~Kharzeev, C.~J.~Na{\"\i}m and Z.~Tu,
``Entanglement as a Probe of Hadronization,''
Phys. Rev. Lett. \textbf{134}, no.11, 111902 (2025)
doi:10.1103/PhysRevLett.134.111902
[arXiv:2410.22331 [hep-ph]].

\bibitem{Levin:2024wtl}
E.~Levin,
``Particle production in a toy model: Multiplicity distribution and entropy,''
Phys. Rev. D \textbf{111}, no.1, 016019 (2025)
doi:10.1103/PhysRevD.111.016019
[arXiv:2412.02504 [hep-ph]].

\bibitem{Ramos:2025tge}
G.~S.~Ramos, L.~S.~Moriggi and M.~V.~T.~Machado,
``Investigating QCD dynamical entropy in high-energy nuclear collisions,''
Phys. Lett. B \textbf{868}, 139737 (2025)
doi:10.1016/j.physletb.2025.139737
[arXiv:2507.09349 [hep-ph]].

\bibitem{Grieninger:2025wxg}
S.~Grieninger, K.~Hao, D.~E.~Kharzeev and V.~Korepin,
``Small $x$ behavior in QCD from maximal entanglement and conformal invariance,''
[arXiv:2508.21643 [hep-ph]].

\bibitem{Kutak:2025awi}
K.~Kutak and M.~Prasza{\l}owicz,
``Cascades of gluons at high energies and their QI measures,''
[arXiv:2511.17288 [hep-ph]].

\bibitem{Sheikhi:2025hep}
S.~Sheikhi and G.~R.~Boroun,
``Entropy and DIS structure functions,''
[arXiv:2511.18285 [hep-ph]].

\bibitem{Cervera-Lierta:2017tdt}
A.~Cervera-Lierta, J.~I.~Latorre, J.~Rojo and L.~Rottoli,
``Maximal Entanglement in High Energy Physics,''
SciPost Phys. \textbf{3}, 036 (2017)
doi:10.21468/SciPostPhys.3.5.036
[arXiv:1703.02989 [hep-th]].

\bibitem{Beane:2018oxh}
S.~R.~Beane, D.~B.~Kaplan, N.~Klco and M.~J.~Savage,
``Entanglement Suppression and Emergent Symmetries of Strong Interactions,''
Phys. Rev. Lett. \textbf{122}, no.10, 102001 (2019)
doi:10.1103/PhysRevLett.122.102001
[arXiv:1812.03138 [nucl-th]].

\bibitem{Hentschinski:2022rsa}
M.~Hentschinski, K.~Kutak and R.~Straka,
``Maximally entangled proton and charged hadron multiplicity in Deep Inelastic Scattering,''
Eur. Phys. J. C \textbf{82}, no.12, 1147 (2022)
doi:10.1140/epjc/s10052-022-11122-1
[arXiv:2207.09430 [hep-ph]].

\bibitem{Asadi:2022vbl}
P.~Asadi and V.~Vaidya,
``Quantum entanglement and the thermal hadron,''
Phys. Rev. D \textbf{107}, no.5, 054028 (2023)
doi:10.1103/PhysRevD.107.054028
[arXiv:2211.14333 [nucl-th]].

\bibitem{Asadi:2023bat}
P.~Asadi and V.~Vaidya,
``1+1D hadrons minimize their biparton Renyi free energy,''
Phys. Rev. D \textbf{108}, no.1, 014036 (2023)
doi:10.1103/PhysRevD.108.014036
[arXiv:2301.03611 [hep-th]].

\bibitem{Hentschinski:2023izh}
M.~Hentschinski, D.~E.~Kharzeev, K.~Kutak and Z.~Tu,
``Probing the Onset of Maximal Entanglement inside the Proton in Diffractive Deep Inelastic Scattering,''
Phys. Rev. Lett. \textbf{131}, no.24, 241901 (2023)
doi:10.1103/PhysRevLett.131.241901
[arXiv:2305.03069 [hep-ph]].

\bibitem{Hatta:2024lbw}
Y.~Hatta and J.~Montgomery,
``Maximally entangled gluons for any x,''
Phys. Rev. D \textbf{111}, no.1, 014024 (2025)
doi:10.1103/PhysRevD.111.014024
[arXiv:2410.16082 [hep-ph]].

\bibitem{Li:2014kfa}
Y.~Li, V.~A.~Karmanov, P.~Maris and J.~P.~Vary,
``Non-Perturbative Calculation of the Scalar Yukawa Theory in Four-Body Truncation,''
Few Body Syst. \textbf{56}, no.6-9, 495-501 (2015)
doi:10.1007/s00601-015-0965-0
[arXiv:1411.1707 [nucl-th]].

\bibitem{Li:2015iaw}
Y.~Li, V.~A.~Karmanov, P.~Maris and J.~P.~Vary,
``Ab Initio Approach to the Non-Perturbative Scalar Yukawa Model,''
Phys. Lett. B \textbf{748}, 278-283 (2015)
doi:10.1016/j.physletb.2015.07.014
[arXiv:1504.05233 [nucl-th]].

\bibitem{Li:2015zah}
Y.~Li,
``Ab initio approach to quantum field theories on the light front,''

\bibitem{Karmanov:2016yzu}
V.~A.~Karmanov, Y.~Li, A.~V.~Smirnov and J.~P.~Vary,
``Nonperturbative solution of scalar Yukawa model in two- and three-body Fock space truncations,''
Phys. Rev. D \textbf{94}, no.9, 096008 (2016)
doi:10.1103/PhysRevD.94.096008
[arXiv:1610.03559 [hep-th]].

\bibitem{Hiller:2016itl}
J.~R.~Hiller,
``Nonperturbative light-front Hamiltonian methods,''
Prog. Part. Nucl. Phys. \textbf{90}, 75-124 (2016)
doi:10.1016/j.ppnp.2016.06.002
[arXiv:1606.08348 [hep-ph]].

\bibitem{Brodsky:1997de}
S.~J.~Brodsky, H.~C.~Pauli and S.~S.~Pinsky,
``Quantum chromodynamics and other field theories on the light cone,''
Phys. Rept. \textbf{301}, 299-486 (1998)
doi:10.1016/S0370-1573(97)00089-6
[arXiv:hep-ph/9705477 [hep-ph]].

\bibitem{Carbonell:1998rj}
J.~Carbonell, B.~Desplanques, V.~A.~Karmanov and J.~F.~Mathiot,
``Explicitly covariant light front dynamics and relativistic few body systems,''
Phys. Rept. \textbf{300}, 215-347 (1998)
doi:10.1016/S0370-1573(97)00090-2
[arXiv:nucl-th/9804029 [nucl-th]].

\bibitem{Bakker:2013cea}
B.~L.~G.~Bakker, A.~Bassetto, S.~J.~Brodsky, W.~Broniowski, S.~Dalley, T.~Frederico, S.~D.~Glazek, J.~R.~Hiller, C.~R.~Ji and V.~Karmanov, \textit{et al.}
``Light-Front Quantum Chromodynamics: A framework for the analysis of hadron physics,''
Nucl. Phys. B Proc. Suppl. \textbf{251-252}, 165-174 (2014)
doi:10.1016/j.nuclphysbps.2014.05.004
[arXiv:1309.6333 [hep-ph]].

\bibitem{Brodsky:2022fqy}
S.~J.~Brodsky, A.~Deur and C.~D.~Roberts,
``Artificial dynamical effects in quantum field theory,''
Nature Rev. Phys. \textbf{4}, no.7, 489-495 (2022)
doi:10.1038/s42254-022-00453-3
[arXiv:2202.06051 [hep-ph]].

\bibitem{Heinzl:2000ht}
T.~Heinzl,
``Light cone quantization: Foundations and applications,''
Lect. Notes Phys. \textbf{572}, 55-142 (2001)
doi:10.1007/3-540-45114-5{\_}2
[arXiv:hep-th/0008096 [hep-th]].

\bibitem{Miller:2000kv}
G.~A.~Miller,
``Light front quantization: A Technique for relativistic and realistic nuclear physics,''
Prog. Part. Nucl. Phys. \textbf{45}, 83-155 (2000)
doi:10.1016/S0146-6410(00)00103-4
[arXiv:nucl-th/0002059 [nucl-th]].

\bibitem{Vary:2009gt}
J.~P.~Vary, H.~Honkanen, J.~Li, P.~Maris, S.~J.~Brodsky, A.~Harindranath, G.~F.~de Teramond, P.~Sternberg, E.~G.~Ng and C.~Yang,
``Hamiltonian light-front field theory in a basis function approach,''
Phys. Rev. C \textbf{81}, 035205 (2010)
doi:10.1103/PhysRevC.81.035205
[arXiv:0905.1411 [nucl-th]].

\bibitem{Alterman:2025prb}
S.~Alterman and P.~J.~Love,
``Entanglement and magic on the light-front,''
[arXiv:2507.10777 [quant-ph]].

\bibitem{Duan:2024dhy}
Y.~Duan, S.~Xu, S.~Cheng, X.~Zhao, Y.~Li and J.~P.~Vary,
``Flavor asymmetry from the nonperturbative nucleon sea,''
Phys. Rev. C \textbf{110}, no.6, 065201 (2024)
doi:10.1103/PhysRevC.110.065201
[arXiv:2404.07755 [hep-ph]].

\bibitem{Horodecki:2009zz}
R.~Horodecki, P.~Horodecki, M.~Horodecki and K.~Horodecki,
``Quantum entanglement,''
Rev. Mod. Phys. \textbf{81}, 865-942 (2009)
doi:10.1103/RevModPhys.81.865
[arXiv:quant-ph/0702225 [quant-ph]].

\bibitem{Guhne:2008qic}
O.~G{\"u}hne and G.~T{\'o}th,
``Entanglement detection,''
Phys. Rept. \textbf{474}, 1-75 (2009)
doi:10.1016/j.physrep.2009.02.004
[arXiv:0811.2803 [quant-ph]].

\bibitem{Cheng:2025zaw}
K.~Cheng, T.~Han and S.~Trifinopoulos,
``Quantum Information at the Electron-Ion Collider,''
[arXiv:2510.23773 [hep-ph]].

\bibitem{Afik:2025ejh}
Y.~Afik, F.~Fabbri, M.~Low, L.~Marzola, J.~A.~Aguilar-Saavedra, M.~M.~Altakach, N.~A.~Asbah, Y.~Bai, H.~Banks and A.~J.~Barr, \textit{et al.}
``Quantum information meets high-energy physics: input to the update of the European strategy for particle physics,''
Eur. Phys. J. Plus \textbf{140}, no.9, 855 (2025)
doi:10.1140/epjp/s13360-025-06752-9
[arXiv:2504.00086 [hep-ph]].

\bibitem{Qi:2025onf}
W.~Qi, Z.~Guo and B.~W.~Xiao,
``Studying Maximal Entanglement and Bell Nonlocality at an Electron-Ion Collider,''
[arXiv:2506.12889 [hep-ph]].

\bibitem{Islam:2015mom}
R.~Islam, R.~Ma, P.~M.~Preiss, M.~E.~Tai, A.~Lukin, M.~Rispoli and M.~Greiner,
``Measuring entanglement entropy through the interference of quantum many-body twins,''
doi:10.1038/nature15750
[arXiv:1509.01160 [cond-mat.quant-gas]].

\bibitem{Hastings:2010zka}
M.~B.~Hastings, I.~Gonz{\'a}lez, A.~B.~Kallin and R.~G.~Melko,
``Measuring Renyi Entanglement Entropy in Quantum Monte Carlo Simulations,''
Phys. Rev. Lett. \textbf{104}, no.15, 157201 (2010)
doi:10.1103/PhysRevLett.104.157201
[arXiv:1001.2335 [cond-mat.str-el]].


\bibitem{Hill:1997pfa}
S.~Hill and W.~K.~Wootters,
``Entanglement of a pair of quantum bits,''
Phys. Rev. Lett. \textbf{78}, 5022-5025 (1997)
doi:10.1103/PhysRevLett.78.5022
[arXiv:quant-ph/9703041 [quant-ph]].

\bibitem{Wootters:1997id}
W.~K.~Wootters,
``Entanglement of formation of an arbitrary state of two qubits,''
Phys. Rev. Lett. \textbf{80}, 2245-2248 (1998)
doi:10.1103/PhysRevLett.80.2245
[arXiv:quant-ph/9709029 [quant-ph]].


\bibitem{Artiaco:2024noa}
C.~Artiaco, T.~Klein Kvorning, D.~Aceituno Ch{\'a}vez, L.~Herviou and J.~H.~Bardarson,
``Universal Characterization of Quantum Many-Body States through Local Information,''
Phys. Rev. Lett. \textbf{134}, no.19, 190401 (2025)
doi:10.1103/PhysRevLett.134.190401
[arXiv:2410.10971 [quant-ph]].

\bibitem{Barata:2025rjb}
J.~Barata, J.~Hormaza, Z.~B.~Kang and W.~Qian,
``Hadronic scattering in (1+1)D SU(2) lattice gauge theory from tensor networks,''
[arXiv:2511.00154 [hep-lat]].

\bibitem{Artiaco:2025qqq}
C.~Artiaco, J.~Barata and E.~Rico,
``Out-of-Equilibrium Dynamics in a U(1) Lattice Gauge Theory via Local Information Flows: Scattering and String Breaking,''
[arXiv:2510.16101 [quant-ph]].

\bibitem{Eisert:2008ur}
J.~Eisert, M.~Cramer and M.~B.~Plenio,
``Area laws for the entanglement entropy - a review,''
Rev. Mod. Phys. \textbf{82}, 277-306 (2010)
doi:10.1103/RevModPhys.82.277
[arXiv:0808.3773 [quant-ph]].

\bibitem{Hentschinski:2024gaa}
M.~Hentschinski, D.~E.~Kharzeev, K.~Kutak and Z.~Tu,
``QCD evolution of entanglement entropy,''
Rept. Prog. Phys. \textbf{87}, no.12, 120501 (2024)
doi:10.1088/1361-6633/ad910b
[arXiv:2408.01259 [hep-ph]].

\bibitem{Brodsky:1998hs}
S.~J.~Brodsky, J.~R.~Hiller and G.~McCartor,
``Pauli-Villars as a nonperturbative ultraviolet regulator in discretized light cone quantization,''
Phys. Rev. D \textbf{58}, 025005 (1998)
doi:10.1103/PhysRevD.58.025005
[arXiv:hep-th/9802120 [hep-th]].

\bibitem{Karmanov:2008br}
V.~A.~Karmanov, J.~F.~Mathiot and A.~V.~Smirnov,
``Systematic renormalization scheme in light-front dynamics with Fock space truncation,''
Phys. Rev. D \textbf{77}, 085028 (2008)
doi:10.1103/PhysRevD.77.085028
[arXiv:0801.4507 [hep-th]].

\bibitem{Zhang:2025wli}
W.~Zhang, Y.~Li and J.~P.~Vary,
``Covariant analysis of electromagnetic current on the light cone: exposition with scalar Yukawa theory,''
JHEP \textbf{07}, 194 (2025)
doi:10.1007/JHEP07(2025)194
[arXiv:2501.07826 [hep-ph]].

\bibitem{Boussarie:2023izj}
R.~Boussarie, M.~Burkardt, M.~Constantinou, W.~Detmold, M.~Ebert, M.~Engelhardt, S.~Fleming, L.~Gamberg, X.~Ji and Z.~B.~Kang, \textit{et al.}
``TMD Handbook,''
[arXiv:2304.03302 [hep-ph]].

\bibitem{Ehlers:2022oal}
P.~J.~Ehlers,
``Entanglement between valence and sea quarks in hadrons of 1+1 dimensional QCD,''
Annals Phys. \textbf{452}, 169290 (2023)
doi:10.1016/j.aop.2023.169290
[arXiv:2209.09867 [hep-ph]].

\bibitem{Pauli:1985ps}
H.~C.~Pauli and S.~J.~Brodsky,
``Discretized Light Cone Quantization: Solution to a Field Theory in One Space One Time Dimensions,''
Phys. Rev. D \textbf{32}, 2001 (1985)
doi:10.1103/PhysRevD.32.2001

\bibitem{Pauli:1985pv}
H.~C.~Pauli and S.~J.~Brodsky,
``Solving Field Theory in One Space One Time Dimension,''
Phys. Rev. D \textbf{32}, 1993 (1985)
doi:10.1103/PhysRevD.32.1993

\bibitem{Bloss:2025ywh}
H.~Bloss, B.~Kriesten and T.~J.~Hobbs,
``Quantum entropy as a harbinger of factorizability,''
Phys. Lett. B \textbf{868}, 139814 (2025)
doi:10.1016/j.physletb.2025.139814
[arXiv:2503.15603 [hep-ph]].

\bibitem{Neill:2018uqw}
D.~Neill and W.~J.~Waalewijn,
``Entropy of a Jet,''
Phys. Rev. Lett. \textbf{123}, no.14, 142001 (2019)
doi:10.1103/PhysRevLett.123.142001
[arXiv:1811.01021 [hep-ph]].

\bibitem{Brodsky:2014yha}
S.~J.~Brodsky, G.~F.~de Teramond, H.~G.~Dosch and J.~Erlich,
``Light-Front Holographic QCD and Emerging Confinement,''
Phys. Rept. \textbf{584}, 1-105 (2015)
doi:10.1016/j.physrep.2015.05.001
[arXiv:1407.8131 [hep-ph]].

\bibitem{Miller:2019ysh}
G.~A.~Miller and S.~J.~Brodsky,
``Frame-independent spatial coordinate $\tilde{z}$: Implications for light-front wave functions, deep inelastic scattering, light-front holography, and lattice QCD calculations,''
Phys. Rev. C \textbf{102}, no.2, 022201 (2020)
doi:10.1103/PhysRevC.102.022201
[arXiv:1912.08911 [hep-ph]].

\bibitem{Angeles-Martinez:2015sea}
R.~Angeles-Martinez, A.~Bacchetta, I.~I.~Balitsky, D.~Boer, M.~Boglione, R.~Boussarie, F.~A.~Ceccopieri, I.~O.~Cherednikov, P.~Connor and M.~G.~Echevarria, \textit{et al.}
``Transverse Momentum Dependent (TMD) parton distribution functions: status and prospects,''
Acta Phys. Polon. B \textbf{46}, no.12, 2501-2534 (2015)
doi:10.5506/APhysPolB.46.2501
[arXiv:1507.05267 [hep-ph]].

\bibitem{Bauer:2001yt}
C.~W.~Bauer, D.~Pirjol and I.~W.~Stewart,
``Soft collinear factorization in effective field theory,''
Phys. Rev. D \textbf{65}, 054022 (2002)
doi:10.1103/PhysRevD.65.054022
[arXiv:hep-ph/0109045 [hep-ph]].

\bibitem{Cloet:2019wre}
I.~C.~Clo{\"e}t, M.~R.~Dietrich, J.~Arrington, A.~Bazavov, M.~Bishof, A.~Freese, A.~V.~Gorshkov, A.~Grassellino, K.~Hafidi and Z.~Jacob, \textit{et al.}
``Opportunities for Nuclear Physics {\&} Quantum Information Science,''
[arXiv:1903.05453 [nucl-th]].

\bibitem{Moriggi:2025qfs}
L.~S.~Moriggi, F.~S.~Navarra and M.~V.~T.~Machado,
``Universality of scaling entropy in charged hadron multiplicity distributions at the LHC,''
Phys. Rev. D \textbf{112}, no.7, 074019 (2025)
doi:10.1103/lynp-yzk5
[arXiv:2506.09899 [hep-ph]].

\bibitem{Brandenburg:2025one}
J.~D.~Brandenburg, S.~R.~Klein, Z.~Xu, S.~Yang, W.~Zha and J.~Zhou,
``Probing quantum phenomena through photoproduction in relativistic heavy-ion collisions,''
Prog. Part. Nucl. Phys. \textbf{143}, 104174 (2025)
doi:10.1016/j.ppnp.2025.104174
[arXiv:2504.18342 [nucl-ex]].

\bibitem{Wang:2022noa}
R.~Wang,
``Classical information entropy of parton distribution functions and an application in searching gluon saturation,''
Eur. Phys. J. A \textbf{60}, no.5, 102 (2024)
doi:10.1140/epja/s10050-024-01322-6
[arXiv:2208.13151 [hep-ph]].

\bibitem{Qian:2021jxp}
W.~Qian, R.~Basili, S.~Pal, G.~Luecke and J.~P.~Vary,
``Solving hadron structures using the basis light-front quantization approach on quantum computers,''
Phys. Rev. Res. \textbf{4}, no.4, 043193 (2022)
doi:10.1103/PhysRevResearch.4.043193
[arXiv:2112.01927 [quant-ph]].

\bibitem{Kreshchuk:2020aiq}
M.~Kreshchuk, S.~Jia, W.~M.~Kirby, G.~Goldstein, J.~P.~Vary and P.~J.~Love,
``Simulating Hadronic Physics on NISQ devices using Basis Light-Front Quantization,''
Phys. Rev. A \textbf{103}, no.6, 062601 (2021)
doi:10.1103/PhysRevA.103.062601
[arXiv:2011.13443 [quant-ph]].

\bibitem{Belyansky:2023rgh}
R.~Belyansky, S.~Whitsitt, N.~Mueller, A.~Fahimniya, E.~R.~Bennewitz, Z.~Davoudi and A.~V.~Gorshkov,
``High-Energy Collision of Quarks and Mesons in the Schwinger Model: From Tensor Networks to Circuit QED,''
Phys. Rev. Lett. \textbf{132}, no.9, 091903 (2024)
doi:10.1103/PhysRevLett.132.091903
[arXiv:2307.02522 [quant-ph]].

\end{thebibliography}
\end{document}